\documentclass[aps,prb,twocolumn,showpacs]{revtex4}

\usepackage{graphicx}

\usepackage{amsmath} \newcommand{\EqLabel}[1]{\label{#1}}
\newcommand{\mb}[1]{\mathbf{#1}}

\begin{document}
 
\title{Systematic improvement of the Momentum Average approximation
  for the Green's function of a Holstein polaron}

\author{Mona Berciu and Glen L. Goodvin}

\affiliation{Department of Physics and Astronomy, University of
  British Columbia, Vancouver, BC V6T 1Z1, Canada }

\date{\today}
 
\begin{abstract}
We show how to systematically improve the Momentum Average (MA)
approximation for the Green's function of a Holstein polaron, by
systematically improving the accuracy of the self-energy diagrams in
such a way that they can still all be summed efficiently. This allows
us to fix some of the problems of the MA approximation, {\em e.g.} we now
find the expected polaron+phonon continuum at the correct location,
and a momentum-dependent self-energy. The quantitative agreement with
numerical data is further improved, as expected since the number of exactly
satisfied spectral weight sum rules is increased. The corrections are
found to be larger in lower dimensional systems.
\end{abstract}

\pacs{71.38.-k, 72.10.Di, 63.20.Kr}

\maketitle

\section{Introduction}

One of the main challenges in condensed matter physics is to find
simple yet accurate analytical approximations  for systems
in non-perturbative (strongly coupled) regimes. This is needed for 
interpretation of experiments and to help supplement the results
available from numerical simulations of model Hamiltonians.

In this general context, we have recently proposed the so-called
Momentum Average (MA) approximation for the Green's function of a
single Holstein polaron.\cite{MA1, MA2} The essence of this
approximation is to sum all diagrams contributing to the polaron
self-energy, however each diagram is approximated to such a degree as to
allow the analytical summation of the entire series. Specifically,
each free propagator appearing in a self-energy diagram is replaced by
its momentum average. The resulting MA self-energy is a
trivial-to-evaluate continued fraction which gives remarkably
accurate results over most of the parameter space, including 
intermediate electron-phonon coupling strengths where perturbational
methods completely fail to capture the correct physics.

In Refs. \onlinecite{MA1} and \onlinecite{MA2} we identified some of
the reasons for this good agreement: first, the MA approximation
becomes asymptotically exact for both very weak and very strong
couplings. More importantly, the resulting MA spectral weight
satisfies the first six spectral weight sum rules exactly, and 
remains highly accurate for all higher order sum rules. However, we
also pointed out some shortcomings of the MA approximation: (i) it
fails to predict the continuum that must appear at $E_{GS} + \Omega$,
where $E_{GS}$ is the polaron ground-state (GS) energy, and $\Omega$
is the frequency of the Einstein phonons (we set $\hbar =1$ throughout
this work). This continuum arises from states that have a phonon
excited very far from where the polaron is. As a result, their
interactions are negligible and the energy of the system is simply the
sum of the two. MA either predicts a wrong location for this continuum
(at weak electron-phonon couplings) or no continuum at all in that
range of energies (at intermediary and strong electron-phonon
couplings).  We noted in Ref. \onlinecite{MA2} that numerical
simulations show that there is very little spectral weight in this
continuum, hence its absence or wrong positioning does not
significantly upset the agreement with the sum rules. Nevertheless, it
would be reassuring to have an approximation that correctly predicts
its existence; (ii) the accuracy of the MA approximation worsens as
$\Omega \rightarrow 0$; (iii) the MA self-energy $\Sigma_{MA}(\omega)$
is independent of the momentum $\mb{k}$ of the electron. Given how
featureless the Holstein model is (electron-phonon coupling and phonon
frequency are both constants) one may expect a rather weak momentum
dependence of the self-energy, however it is certainly not entirely
absent.

In this article, we show how to systematically improve the MA
approximation, generating a hierarchy of approximations that we call
MA$^{(n)}$ (the original MA is MA$^{(0)}$ in this notation). As
explained below, the idea is to systematically improve the accuracy of
the ``simplified'' self-energy diagrams. The results become more and
more accurate as $n$ increases -- for example, while the MA spectral
weight satisfies only the first 6 sum rules exactly, this improves to
8 and 10 exact sum rules respectively for MA$^{(1)}$ and MA$^{(2)}$ spectral
weights. While  the numerical effort also
increases, it is still trivial for the $n=1$ and $n=2$ levels
that we discuss explicitly here. Level MA$^{(1)}$ already solves the
continuum problem, while all levels with $n \ge 2$ produce
momentum-dependent self-energies. The accuracy in the limit $\Omega
\rightarrow 0$ is also shown to improve significantly with increasing
$n$. In effect, for a slightly increased numerical effort, MA$^{(2)}$
solves all the known problems of the MA$^{(0)}$ approximation.

The work is organized as follows: in Sec. II we briefly review the
MA$^{(0)}$ approximation, presenting a new argument to explain its
accuracy. In Sec. III we describe the systematic approach to obtain
the improved versions MA$^{(n)}$, $n\ge 1$, and give explicit formulae
for the self-energies corresponding to the MA$^{(1)}$ and MA$^{(2)}$
approximations. In Sec. IV we compare the predictions of these
approximations against numerical simulations, to gauge the improved
accuracy as $n$ increases. Spectral sum rules, as well as variational
arguments, will also be used to explain the systematic improvement of
accuracy with increasing $n$. Finally, Sec. V contains our conclusions.

\section{Brief review of the MA$^{(0)}$ approximation}

The Holstein model is the simplest lattice model that includes
electron-phonon coupling. Its Hamiltonian is:\cite{Holstein}
\begin{equation}
\nonumber {\cal H}=\sum_{\mb{k}} \left( \varepsilon_{\mb{k}} c_{\mb{k}
}^{\dagger} c_{\mb{k}}+\Omega b_{\mb{k}}^{\dagger}b_{\mb{k}} \right)
\\ + \frac{g}{\sqrt{N}} \sum_{\mb{k}, \mb{q}}
c_{\mb{k}-\mb{q}}^{\dagger} c_{\mb{k}} \left(b_{\mb{q}}^{\dagger} +
b_{\mb{-q}} \right).
\end{equation}
The first term is the kinetic energy of the electron, with
$c_{\mb{k}}^{\dagger}$ and $c_{\mb{k}}$ being the electron creation
and annihilation operators. For the single electron (polaron) problem
of interest to us, the spin of the electron is irrelevant and we
suppress its index.  $\varepsilon_{\mb{k}}$ is the free-particle
dispersion. In all results shown here, we assume nearest-neighbor
hopping on a $d$-dimensional simple cubic lattice of constant $a$ (we
set $a=1$) with a total of $N$ sites, and with periodic boundary
conditions. In this case
\begin{equation}
\nonumber \varepsilon_{\mb{k}}=-2t \sum_{i=1}^d \cos(k_i a),
\end{equation}
but our results are valid for any other dispersion.  The second term
describes a branch of optical phonons of energy $\Omega$.
$b_{\mb{q}}^{\dagger}$ and $b_{\mb{q}}$ are the phonon creation and
annihilation operators. The last term is the on-site linear
electron-phonon coupling $V_{{\rm el-ph}}=g\sum_{i}^{} c^\dag_i c_i
(b^\dag_i + b_i)$, written in $\mb{k}$-space. All sums over momenta
are over the first Brillouin zone, namely $-{\pi} < k_i \leq {\pi}$,
$i=1,d$.

The quantity of interest to us is the (retarded) single polaron
Green's function, defined as:\cite{MA1,MA2}
\begin{equation}
\EqLabel{3} G({\mb k}, \omega) = \langle 0 | c_{\mb k} \hat{G}(\omega)
c_{\mb k}^\dag|0\rangle = \langle 0 | c_{\mb k} {1\over \omega - {\cal
H} + i \eta}c_{\mb k}^\dag|0\rangle
\end{equation}
where $|0\rangle$ is the vacuum $c_{\mb k}|0\rangle = b_{\mb
  q}|0\rangle =0$, and $\eta > 0 $ is infinitesimally small.

As described in detail in Ref. \onlinecite{MA2}, using repeatedly
Dyson's identity $\hat{G}(\omega)=\hat{G}_0(\omega)+
\hat{G}(\omega)V_{{\rm el-ph}}\hat{G}_0(\omega)$, where $V_{{\rm
el-ph}} = {\cal H} - {\cal H}_0$ is the electron-phonon interaction
potential and $\hat{G}_0(\omega) =\left[\omega - {\cal H}_0 + i
\eta\right]^{-1}$, we generate the infinite hierarchy of equations of
motion whose exact solution is the desired Green's function:
\begin{equation} \label{eq:G}
G(\mb{k},\omega) = G_0(\mb{k}, \omega) \left[
1+\frac{g}{\sqrt{N}}\sum_{\mb{q}_1} F_1(\mb{k},\mb{q}_1,\omega)
\right],
\end{equation}
and for $n \ge 1$,
\begin{widetext}
\begin{equation} \label{eq:F}
F_n(\mb{k}, \mb{q}_1, \ldots, \mb{q}_n, \omega) = \frac{g}{\sqrt{N}}
G_0(\mb{k}-\mb{q}_T,\omega-n\Omega) \left[ \sum_{i=1}^{n}
F_{n-1}(\mb{k}, \ldots, \mb{q}_{i-1}, \mb{q}_{i+1}, \ldots, \omega)
\right. + \left. \sum_{\mb{q}_{n+1}} F_{n+1}(\mb{k}, \mb{q}_1, \ldots,
\mb{q}_{n+1}, \omega) \right].
\end{equation}
\end{widetext}
Here, ${\mb q_T} = \sum_{i=1}^{n} {\mb q}_i$ is the total momentum
carried by phonons, $G_0({\mb k}, \omega) = (\omega - \epsilon_{\mb k}
+ i \eta)^{-1}$ is the free electron Green's function, and we
introduced the generalized Green's functions
\begin{equation}
\nonumber F_n(\mb{k},\mb{q}_1,\dots,\mb{q}_n,\omega)=\langle0|
c_{\mb{k}} \hat{G}(\omega)c_{\mb{k}-\mb{q}_T}^{\dagger}
b_{\mb{q}_1}^{\dagger}\dots b_{\mb{q}_n}^{\dagger}|0 \rangle.
\end{equation}

These equations can be recast in a more convenient form after
observing that if we treat Eqs. (\ref{eq:F}) as an inhomogeneous
system of linear equations in unknowns $F_1, F_2, ...$, then the only
inhomogeneous term appears in the first equation and is proportional
to $F_0 ({\mb k}, \omega) = G({\mb k}, \omega)$. It follows that all
generalized Green's functions $F_1, F_2, ...$ must be proportional to
$G({\mb k}, \omega)$. As a result we introduce the more convenient
variables:
\begin{equation}
\EqLabel{5} f_n({\mb k}, \mb{q}_1, \ldots, \mb{q}_n, \omega) =
\frac{N^{n\over 2} g^n F_n(\mb{k}, \mb{q}_1, \ldots, \mb{q}_n, \omega)
}{G({\mb k}, \omega)}.
\end{equation}
In terms of these, Eq. (\ref{eq:G}) becomes:
\begin{equation}
\nonumber G(\mb{k},\omega) = G_0(\mb{k}, \omega) \left[
1+\frac{1}{N}\sum_{\mb{q}_1} f_1(\mb{k},\mb{q}_1,\omega) G({\mb k},
\omega)\right]
\end{equation}
so that the exact self-energy is:
\begin{equation}
\EqLabel{10} \Sigma({\mb k}, \omega) = \frac{1}{N}\sum_{\mb{q}_1}
f_1(\mb{k},\mb{q}_1,\omega),
\end{equation}
giving the standard solution:
\begin{equation}
\EqLabel{10a} G({\mb k}, \omega) =\frac{1}{\omega-\epsilon_{\bf k} -
\Sigma({\mb k}, \omega) +i\eta}.
\end{equation}
To find $ f_1(\mb{k},\mb{q}_1,\omega)$, we must solve the infinite
system of coupled equations that result from Eqs. (\ref{eq:F}). For
later convenience, we write the first few equations explicitly here,
using the short-hand notation $f_n(\mb{k}, \mb{q}_1, \ldots, \mb{q}_n,
\omega) \equiv f_n(\mb{q}_1, \ldots, \mb{q}_n)$ (i.e., the dependence
of ${\mb k}$ and $\omega$ of these functions is implicitly assumed
from now on). Then:
\begin{widetext}
\begin{eqnarray}
 \EqLabel{e1} & &f_1(\mb{q}_1) = G_0(\mb{k}-{\mb q}_1, \omega-\Omega
)\left[g^2 + {1\over N} \sum_{{\mb q}_2}^{} f_2({\mb q}_1, {\mb
q}_2)\right], \\ \EqLabel{e2} & &f_2( \mb{q}_1, {\mb q}_2) =
G_0(\mb{k}-{\mb q}_1-{\mb q}_2,\omega-2\Omega)\left[g^2
\left[f_1(\mb{q}_1) + f_1(\mb{q}_2)\right] + {1\over N} \sum_{{\mb
q}_3}^{} f_3( {\mb q}_1, {\mb q}_2, {\mb q}_3)\right]
\end{eqnarray}
and for all $n\ge 3$,
\begin{equation} \label{e3}
f_n(\mb{q}_1, \ldots, \mb{q}_n) =
G_0(\mb{k}-\sum_{i=1}^{n}\mb{q}_i,\omega-n\Omega) \left[
g^2\sum_{i=1}^{n} f_{n-1}(\mb{q}_1,\ldots, \mb{q}_{i-1}, \mb{q}_{i+1},
\ldots, \mb{q}_n) + {1\over N} \sum_{\mb{q}_{n+1}} f_{n+1}(\mb{q}_1,
\ldots, \mb{q}_{n+1}) \right].
\end{equation}
\end{widetext}
Clearly, all the dependence on free propagators of the self-energy
comes from the free propagator prefactors on the right-hand side of
these equations.

As already stated, MA$^{(0)}$ consists in replacing all free
propagators in all self-energy diagrams by their momentum average:
\begin{equation}
\label{eq:g0} \bar{g}_0(\omega)=\frac{1}{N}\sum_{\mb{k}}
G_0(\mb{k},\omega).
\end{equation}
Obviously, this corresponds to replacing the free propagator
pre-factors on the right-hand side of Eqs. (\ref{e1})-(\ref{e3}) by
the corresponding $\bar{g}_0(\omega- n\Omega)$. In this case, it is
straightforward to see that the resulting solutions, denoted
$f_n^{(0)}$, are functions of $\omega$ only (all dependence on phonon
momenta disappears at this level of approximation). The resulting
equations $f_1^{(0)}(\omega) = \bar{g}_0(\omega-\Omega)\left[ g^2 +
f_2^{(0)}(\omega)\right]$ and for $n\ge 2$, $f_n^{(0)}(\omega) =
\bar{g}_0(\omega-n\Omega)\left[ n g^2 f_{n-1}^{(0)}(\omega)+
f_{n+1}^{(0)}(\omega)\right]$ are solved in terms of continued
fractions\cite{book} (also see Appendix \ref{ap1}) to find:\cite{MA2}
\begin{equation}
\EqLabel{ma0} \Sigma_{MA^{(0)}}(\omega) = f_1^{(0)}(\omega) = g^2
A_1(\omega)
\end{equation}
where we define the infinite continued fractions:
\begin{eqnarray}
\EqLabel{cf} A_n(\omega) &=& \frac{n \bar{g}_0(\omega-n\Omega)}{1- g^2
\bar{g}_0(\omega-n\Omega) A_{n+1}(\omega)}\\ \nonumber &=&\cfrac{n
\bar{g}_0(\omega-n\Omega)}{1- \cfrac{(n+1)g^2
\bar{g}_0(\omega-n\Omega)\bar{g}_0(\omega-(n+1)\Omega) }{1-\dots }}.
\end{eqnarray}
Results based on this MA$^{(0)}$ approximation have been analyzed in
detail in Ref. \onlinecite{MA2}.

\begin{figure}[t]
\includegraphics[width=0.80\columnwidth]{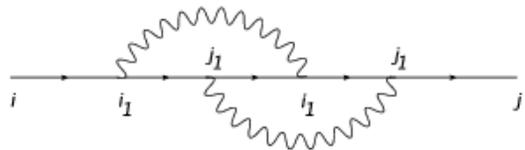}
\caption{A second order diagram contribution to $G(i,j,\omega)$.}
\label{fig1}
\end{figure}

Before discussing how to systematically improve this approximation, it
is worth pointing out an alternative explanation for its good
accuracy.\cite{Bar} This comes from realizing that in real space, the
meaning of the MA$^{(0)}$ approximation is that it replaces all
free-propagators $G_0(i,j,\omega-n\Omega)$ appearing in all
self-energy diagrams, by $\delta_{i,j} \bar{g}_0(\omega-n\Omega)$. For
example, consider the real-space, second-order Green's function
diagram depicted in Fig. \ref{fig1}. It has the exact value $
\sum_{i_1,j_1}^{}G_0(i,i_1,\omega)
\Sigma_{2,c}(i_1,j_1,\omega)G_0(j_1, j,\omega), $ where the
contribution to self-energy from the second-order crossed self-energy
diagram is: $ \Sigma_{2,c}(i_1,j_1,\omega) = g^4
G_0(i_1,j_1,\omega-\Omega) G_0(j_1,i_1,\omega - 2\Omega)
G_0(i_1,j_1,\omega-\Omega).  $ Within MA$^{(0)}$,
$\Sigma_{2,c}(i_1,j_1,\omega)$ is approximated as
$\Sigma^{(0)}_{2,c}(i_1,j_1,\omega)=\delta_{i_1,j_1}
\Sigma^{(0)}_{2,c}(\omega)= \delta_{i_1,j_1}g^4
\bar{g}_0(\omega-\Omega)\bar{g}_0(\omega-2\Omega)\bar{g}_0(\omega-\Omega)$.
Inserting this into the Green's function diagram removes one of the
sums, and after Fourier transforming we find that the contribution of
this diagram to $G({\bf k},\omega)$ is $G_0({\mb k}, \omega)
\Sigma^{(0)}_{2,c}(\omega) G_0({\mb k}, \omega)$. The same holds for
all higher order diagrams.  Summing all of them, we see that the free
propagators in the {\em proper self-energy parts} have indeed been
replaced by their momentum averages.

The reason why it is a good zero-order approximation to keep only the
diagonal (in real space) contributions
$G_0(i,j,\omega-n\Omega)\rightarrow \delta_{i,j}
\bar{g}_0(\omega-n\Omega)$ is straightforward to understand, at least
for low energies $\omega \sim E_{GS}$. Because of interactions,
$E_{GS} < -2dt$ (the polaron ground-state is below the free electron
continuum). It follows that for $\omega \sim E_{GS}$, the free
propagators $G_0(i,j,\omega-n\Omega)$ are needed at energies below the
free electron continuum, and the larger $n$ is the further below the
band-edge these energies are.  However, it is well-known that for
$\omega < - 2dt$, the free propagator decreases exponentially with
increasing distance $|i-j|$. The reason is that there are no
free-electron eigenstates outside the free-electron band, and the
electron has to tunnel from 
one site to another. 
For example, in one dimension\cite{Eco}
$$ G_0(i,j,\omega) = e^{ik_0|i-j|} G_0(i,i,\omega) =e^{ik_0|i-j|}
\bar{g}_0(\omega)
$$ where $k_0$ is the first quadrant solution of the equation $\omega
= - 2t\cos k_0$. For $\omega < - 2t$, $k_0 = i \kappa$, where $\kappa
> 0$ increases as $\omega$ decreases. We also used the fact that in
terms of its Fourier transform:
$$ G_0(i,i,\omega) ={1\over N} \sum_{\mb k}^{} e^{i{\mb k}\cdot({\mb
R_i} - {\mb R_i})} G_0({\mb k},\omega) = \bar{g}_0(\omega).
$$

It follows that at low energies, MA$^{(0)}$ keeps the largest
contributions to the self-energy (from real-space diagonal terms),
while ignoring exponentially small contributions coming from
off-diagonal terms. This is expected to become more and more accurate
for higher order diagrams with many phonons: the larger $n$ is, the
faster the exponential decay with distance of
$G_0(i,j,\omega-n\Omega)$ becomes.

Based on these arguments one expects at least low-energy
properties to be well described by MA, at least if $\Omega$ is not
too small. Together with the good agreement\cite{MA1,MA2} with the
spectral weight sum rules, this leads to the conclusion that the
self-energy and the Green's function at all energies should indeed be
quite accurate within MA. Comparison with numerics validates
this.\cite{MA2}

\section{Higher levels MA$^{(n)}, n\ge 1$}

The above arguments also suggest a systematic way towards improving
the MA$^{(0)}$ approximation. The biggest error at low energies is due
to the momentum average of the propagators of energy
$\omega-\Omega$. This is the energy closest to the free electron
continuum and therefore these propagators have the slowest decay in
real space. The next slowest decay is for the propagators of energy
$\omega-2\Omega$, etc. If one could selectively keep these propagators
exactly while momentum-averaging the ones with more phonons (lower
energy, faster decay) this should improve the accuracy of the
approximation at low energies. In fact, all sum rules would also be
improved (individual diagrams are more accurate) therefore one would
expect an improvement at all energies.  This is precisely what the
higher levels MA$^{(n)}$, $n\ge 1$, achieve.

We define MA$^{(n)}$ as being the approximation where in all
self-energy diagrams, all free propagators with energy $\omega - m
\Omega$, where $m\le n$, are kept exactly, while the ones with more
than $n$ phonons are momentum averaged.

In terms of the equations-of-motion that need to be solved,
Eqs. (\ref{e1})-(\ref{e3}), achieving this is straightforward, since a
propagator of energy $\omega - m \Omega$ appears only once, in the
right-hand side pre-factor of the equation for $f_m({\bf q}_1, \ldots,
{\bf q}_m)$. It follows that if we keep the first $n$ of
Eqs. (\ref{e1})-(\ref{e3})  as they are, and approximate the
equations for $f_{n+1}, f_{n+2}, \ldots$ by momentum-averaging the
free propagator appearing in the right-hand side pre-factor $G_0({\mb
k}-\mb{q}_T, \omega - m \Omega)\rightarrow \bar{g}_0(\omega-m\Omega)$
if $m\ge n+1$, we achieve our goal -- provided that we can find the
solution of the resulting infinite system of coupled equations.

We now derive explicitly the solutions for $n=1$ and $n=2$ levels.

\subsection{MA$^{(1)}$ level}

In this case, the equations to be solved are:
\begin{equation}
\EqLabel{s1} \Sigma_{MA^{(1)}} ({\bf k}, \omega) = {1\over N}
\sum_{{\mb q}_1}^{} f_1^{(1)}({\mb q_1})
\end{equation}
where
\begin{equation}
\EqLabel{e11} f_1^{(1)}(\mb{q}_1) = G_0(\mb{k}-{\mb q_1},
\omega-\Omega )\left[g^2 + {1\over N} \sum_{{\mb q}_2}^{}
f_2^{(1)}({\mb q_1}, {\mb q_2})\right]
\end{equation}
and for all $n\ge 2$,
\begin{widetext}
\begin{equation}
\EqLabel{e21} f_n^{(1)}(\mb{q}_1, \ldots, \mb{q}_n) =
\bar{g}_0(\omega-n\Omega) \left[ g^2\sum_{i=1}^{n}
f^{(1)}_{n-1}(\ldots, \mb{q}_{i-1}, \mb{q}_{i+1}, \ldots) + {1\over N}
\sum_{\mb{q}_{n+1}} f_{n+1}^{(1)}(\mb{q}_1, \ldots, \mb{q}_{n+1})
\right].
\end{equation}
\end{widetext}
As before, the dependence on ${\mb k}, \omega$ is implicitly assumed
everywhere.  We use the upper labels $(1)$ because these are the
approximative solutions corresponding to MA$^{(1)}$.

The solution of this infinite set of recurrent equations is discussed in
Appendix \ref{ap2}. The end result is:
\begin{equation}
\EqLabel{ma1} \Sigma_{MA^{(1)}}(\omega) = \frac{g^2
\bar{g}_0(\tilde{\omega})}{1- g^2
\bar{g}_0(\tilde{\omega})\left[A_2(\omega) -
A_1(\omega-\Omega)\right]},
\end{equation}
where [see Eq. (\ref{ma0})]:
\begin{equation}
\EqLabel{tw} \tilde{\omega} = \omega - \Omega - g^2
A_1(\omega-\Omega)=\omega - \Omega - \Sigma_{MA}(\omega-\Omega) .
\end{equation}
The continued fractions $A_1(\omega-\Omega), A_2(\omega)$ are defined
in Eq.~(\ref{cf}). This expression is slightly more complicated than
$\Sigma_{MA^{(0)}}(\omega)$, since it involves two different continued
fractions, however it is still very trivial to compute.

Note that based on this and other results derived in Appendix
\ref{ap2}, we can now calculate the MA$^{(1)}$ expressions for the
generalized Green's functions $f_n^{(1)}({\mb k}, {\mb q}_1, \ldots,
{\mb q}_n,\omega)\sim F_n({\mb k}, {\mb q}_1, \ldots, {\mb
q}_n,\omega)$. These will be more accurate than the values obtained
within the MA$^{(0)}$ approximation, where none of the $f_n^{(0)}$
expression had any momentum dependence. These generalized Green's
functions contain further information about the polaron, for example regarding
the phonon statistics.

As can be seen from Eq. (\ref{ma1}), the self-energy is still momentum
independent at this level. However, it is clear that this is because
the Holstein model is so featureless. If, {\em e.g.} the coupling was
dependent on the phonon momentum, then the first self-energy diagram
${1\over N} \sum_{{\mb q}}^{} |g({\mb q})|^2 G_0({\mb k} - {\mb q},
\omega)$ would be ${\mb k}$ dependent, and so would
$\Sigma_{MA^{(1)}}({\mb k}, \omega)$ (this diagram is exact at
MA$^{(1)}$ level). Indeed, work in progress on generalizing this
approach to models with $g({\mb q})$ coupling verifies this.

It is also clear that even for the Holstein model, all expressions
$\Sigma_{MA^{(n)}}({\mb k}, \omega)$ with $n\ge 2$ will have momentum
dependence, since Holstein self-energy diagrams of second order are
momentum dependent. We demonstrate below that this is indeed the case.

Finally, the MA approximation becomes exact in the limit $g
\rightarrow 0$ and $t\rightarrow 0$. The first limit is trivial, since
$\Sigma \rightarrow 0$. The second is due to the fact that if $t=0$
then free propagators $G_0({\mb k}, \omega)$ are in fact independent
of ${\mb k}$, and thus the momentum averages become
irrelevant. Clearly, the same must hold true for all higher level
MA$^{(n)}$ approximations.  Indeed, one can verify directly that if
$\bar{g}_0(\omega) = (\omega + i\eta)^{-1}$ (corresponding to $t=0$),
then the expressions in Eqs. (\ref{ma1}) and (\ref{ma0}) are
equal. The same is true for the MA$^{(2)}$ results we present below.

\begin{widetext}

\subsection{MA$^{(2)}$ level}

In this case, the equations to be solved are [compare to the exact
  Eqs. (\ref{e1})-(\ref{e3})]:
\begin{equation}
\EqLabel{s2} \Sigma_{MA^{(2)}} ({\bf k}, \omega) = {1\over N}
\sum_{{\mb q}_1}^{} f_1^{(2)}({\mb q_1})
\end{equation}
where
\begin{eqnarray}
 \EqLabel{e31} & &f_1^{(2)}(\mb{q}_1) = G_0(\mb{k}-{\mb q_1},
\omega-\Omega )\left[g^2 + {1\over N} \sum_{{\mb q}_2}^{}
f_2^{(2)}({\mb q_1}, {\mb q_2})\right], \\ \EqLabel{e22} & &f_2^{(2)}(
\mb{q}_1, {\mb q_2}) = G_0(\mb{k}-{\mb q_1}-{\mb
q_2},\omega-2\Omega)\left[g^2 \left[f_1^{(2)}(\mb{q}_1) +
f_1^{(2)}(\mb{q}_2)\right] + {1\over N} \sum_{{\mb q}_3}^{} f_3^{(2)}(
{\mb q_1}, {\mb q_2}, {\mb q_3})\right]
\end{eqnarray}
and for all $n\ge 3$,
\begin{equation} \label{e32}
f_n^{(2)}(\mb{q}_1, \ldots, \mb{q}_n) = \bar{g}_0(\omega-n\Omega)
\left[ g^2\sum_{i=1}^{n} f^{(2)}_{n-1}(\ldots, \mb{q}_{i-1},
\mb{q}_{i+1}, \ldots) + {1\over N} \sum_{\mb{q}_{n+1}}
f^{(2)}_{n+1}(\mb{q}_1, \ldots, \mb{q}_{n+1}) \right].
\end{equation}
\end{widetext}
Dependence on ${\mb k}, \omega$ is again implicitly assumed.

This can be reduced to a closed system of equations in terms of only
$f_1^{(2)}(\mb{q}_1)$ and $f_2^{(2)}( \mb{q}_1, {\mb q_2})$, after
solving for ${1\over N} \sum_{{\mb q}_3}^{} f_3^{(2)}( {\mb q_1}, {\mb
q_2}, {\mb q_3})$ from Eqs. (\ref{e31}). The details are provided in
Appendix \ref{ap3}.

We use the short-hand notation:
$$ A_1 \equiv A_1(\omega-2\Omega); A_2\equiv A_2(\omega-\Omega); A_3
\equiv A_3(\omega),$$ where the continued fractions are defined in
Eq. (\ref{cf}). We also define various momentum averages (dependence
on ${\mb k}, \omega$ is implicit):
\begin{eqnarray}
\nonumber && {\cal F}_1 = \Sigma_{MA^{(2)}} = {1\over N} \sum_{{\mb
 q}_1}^{} f_1^{(2)}({\mb q_1}), \\ \nonumber && {\cal F}_2 = {1\over
 N^2} \sum_{{\mb q}_1, {\mb q}_2}^{} f_2^{(2)}({\mb q_1}, {\mb q_2}) =
 \frac{2g^2\bar{g}_0(\tilde{\omega}){\cal
 F}_1}{1-g^2\bar{g}_0(\tilde{\omega})(A_3-A_1)}, \\ \nonumber &&\delta
 \bar{f}_2({\mb q}_1) = {1\over N} \sum_{{\mb q}_2}^{} f_2^{(2)}({\mb
 q_1}, {\mb q_2}) -{\cal F}_2.
\end{eqnarray}
The link between ${\cal F}_1$ and ${\cal F}_2$ is proved in
Eq. (\ref{F2}).

In terms of these, the closed system of equations to be solved becomes
(see Appendix \ref{ap3} for more details):
\begin{equation}
\nonumber 
f_1^{(2)}({\mb q_1})=G_0({\mb k} - {\mb q}_1,
\omega -\Omega)\left[g^2 + \delta \bar{f}_2({\mb q}_1) +{\cal F}_2\right],
\end{equation}
\begin{eqnarray}
\nonumber \delta \bar{f}_2({\mb q}_1) = g^2
\bar{g}_0(\tilde{\omega})\left[f_1^{(2)}({\mb q_1}) + (A_2-A_1)\delta
\bar{f}_2({\mb q}_1)- 2{\cal F}_1\right] && \\ \nonumber 
+{g^2\over N} \sum_{{\mb q}_2}^{} G_0({\mb k}-{\mb q}_1 - {\mb q}_2,
\tilde{\omega})\left[f_1^{(2)}({\mb q_2}) + (A_2-A_1)\delta
\bar{f}_2({\mb q}_2)\right].&&
\end{eqnarray}

These equations can be solved in a variety of ways. We present here
the most efficient solution that we have found, and then comment
briefly on other possible solutions. First, given the form of these
equations, it is advantageous to introduce the new unknown
\begin{equation}
\EqLabel{x} x_{\mb q} = f_1^{(2)}({\mb q}) + (A_2-A_1)\delta
\bar{f}_2({\mb q}).
\end{equation}
Consider its Fourier transform at various lattice sites ${\mb R}_i$,
  namely $x(i) = {1\over N}\sum_{\mb q}^{}e^{i{\mb q} \cdot {\mb R}_i}
  x_{\mb q}$. First, observe that $x(0) = {1\over N}\sum_{\mb
  q}^{}x_{\mb q} = {\cal F}_1 = \Sigma_{MA^{(2)}}$, by definition.

As shown in Appendix \ref{ap3}, the set of two closed equations above
can be rewritten as an inhomogeneous equation involving $x(i)$ (at all
lattice sites):
\begin{equation}
\EqLabel{x1} \sum_{j}^{} M_{ij}({\mb k}, \omega) x(i) = e^{i{\mb k}
\cdot {\mb R}_i} g^2 G_0(-i, \tilde{\tilde{\omega}})
\end{equation}
where
$$ \tilde{\tilde{\omega}}= \omega -\Omega -\frac{g^2
\bar{g}_0(\tilde{\omega})}{1- g^2 \bar{g}_0(\tilde{\omega})(A_2-A_1)},
$$ $G_0(i,\omega) = {1\over N}\sum_{\mb k}^{}e^{i{\mb k} \cdot {\mb
R}_i} G_0({\mb k}, \omega), $ and the expression of the matrix
elements $M_{ij}({\mb k},\omega)$ is given in
Eqs. (\ref{moo}-\ref{mij}). They have simple expressions, involving
only (the same) three continued fractions $A_1, A_2, A_3$ as well as
various $G_0(i,\omega)$ values, therefore they can be calculated easily.

Because at low energies the free propagators in real space decay
exponentially, one expects that $x(i)$ decreases fast with increasing
distance ${\mb R}_i$. Alternatively, consider, for instance, the
$f_1^{(2)}({\mb q})$ contribution to $x_{\mb q}$. When Fourier
transformed, the initial state $c_{{\mb k}-{\mb q}}^\dag b_{\mb
q}^\dag|0\rangle$ goes into $c_{j}^\dag b_{j+i}^\dag|0\rangle$, i.e.
the phonon is further and further apart from the electron. Similar
interpretation can be given for the second contribution to $x(i)$. One
expects the amplitudes for such processes to decay with $i$.

As a result, we can truncate the system of equations (\ref{x1}),
assuming that $x(i)=0$ for all ${\mb R}_i$ larger than a cutoff. We
vary this cutoff to insure that convergence has indeed been 
achieved. In this formulation, we find that convergence is reached
extremely fast, typically for a cutoff distance of order $5a$ (see
results section). In other words, to obtain ${ \Sigma}_{MA^{(2)}}({\mb
k}, \omega)=x(0)$, in 1D we typically have to solve a system of $11$
or so inhomogeneous equations, which can be done very
efficiently. Higher dimensions imply larger systems, but overall the
numerical task is still trivial and results can be obtained very fast
and with little computational resources.

It is important to emphasize that such low cutoffs are {\em not}
inherent in the problem. In fact, one could also solve these
equations, for instance, by eliminating $f_1^{(2)}({\mb q})$ to get an
equation only in terms of $\delta \bar{f}_2({\mb q}_1)$ and ${\cal
F}_2 \sim \Sigma$. If one Fourier transforms this, it turns out that
cutoffs as large as $100a$ are needed before convergence is achieved,
and this is especially so for the polaron+one-phonon continuum (the
bound polaron states converge quickly). This is not surprising,
since states in the polaron+one-phonon continuum do have a free phonon,
i.e. one that could be infinitely far from the polaron. In reality, we
expect that if we have a  finite but large enough system, all
quantities will eventually converge to their bulk values. In
particular, here this suggests that one needs to allow the free phonon
to be hundreds of sites away from the polaron before convergence for
the continuum is achieved.

The much faster convergence for the formulation of Eq. (\ref{x1}) is
due in part to the particular choice of variable $x_{\mb q}$. Even
more important is the infinite summation of diagrams performed when
the new frequency $\tilde{\tilde{\omega}}$ appears (see Appendix
\ref{ap3}). Without this, the convergence for continuum energies
remains slow and large cutoffs are needed. Examples are discussed in
the results section.

Of course, one could also attempt to solve these equations directly in
the ${\mb k}$-space. Without having tried it, we believe this to be an
inefficient approach. The goal is to find ${\cal F}_1=\Sigma$, {\em
i.e.}  an average over the Brillouin zone. Within MA$^{(1)}$,
$f_1^{(1)}({\mb q})\sim G_0({\mb k} - {\mb q}, \tilde{\omega})$ (it is
a constant in MA).  Presuming that $f_1^{(2)}({\mb q})$ is not too
different, it is clear that these functions are of
comparable size everywhere in the Brillouin zone, and therefore one
should sample many points in the Brillouin zone to obtain an accurate
 average.

Finally, going back to Eq. (\ref{x1}), we would like to point out that if
we set the cutoff at zero distance, i.e. use $M_{00} x(0) = g^2
\bar{g}_0(\tilde{\tilde{\omega}})$, we obtain an analytical,
momentum-independent approximation to the true $\Sigma_{MA^{(2)}}({\mb
k}, \omega)$:
\begin{equation}
\EqLabel{x2}
\tilde{\Sigma}_{MA^{(2)}}(\omega)= \frac{g^2
\bar{g}_0(\tilde{\tilde{\omega}})}{1- g^2
\bar{g}_0(\tilde{\omega})\bar{g}_0(\tilde{\tilde{\omega}})\left({2\over
a_{31}(\omega)}-{1\over a_{21}(\omega)}\right)},
\end{equation}
where
$$ a_{ij}(\omega) = 1 -g^2 \bar{g}_0(\tilde{\omega})(A_i-A_j).
$$ One can think of this as the variant of MA$^{(2)}$ that keeps some
of the free propagators of energy $\omega-2\Omega$ exactly (typically
those appearing in non-crossed diagrams), but averages over those that
give momentum dependence to the self-energy. The self-energy
$\tilde{\Sigma}_{MA^{(2)}}(\omega)$ is more accurate than
$\Sigma_{MA^{(1)}}(\omega)$ but less accurate than
$\Sigma_{MA^{(2)}}({\mb k}, \omega)$. Such ``zero-cutoff'' analytical
approximations can be obtained for higher levels of MA$^{(n)}$ quite
easily. The full MA$^{(n)}$ for $n > 2$ can also be done. The
reduction to a closed system of $n$ coupled equations is always
straightforward. Its solution, however, becomes more computationally
involved as $n$ increases, and leads to gradually smaller improvements
in the accuracy.

\section{Results}

A detailed comparison of the predictions of the MA approximation
vs. numerical simulations is available in
Ref. \onlinecite{MA2}. Instead of another comprehensive investigation,
here we will focus on several properties where the higher level
MA$^{(n)}$ approximations show a significant improvement over MA
results. The way of extracting quantities of interest from the Green's
function, {\em e.g.} ground-state (GS) energies, quasiparticle ($qp$)
weights, effective masses, average number of phonons in the polaron
cloud, etc., are described in detail in
Ref. \onlinecite{MA2}. Throughout we use
\begin{equation}
\EqLabel{la} \lambda = {g^2\over 2dt \Omega}
\end{equation}
as the effective coupling strength, $d$ being the dimension of the
lattice. All energies are measured in units of $t$.

\subsection{Ground-state properties}

\begin{figure}[t]
\includegraphics[width=0.90\columnwidth]{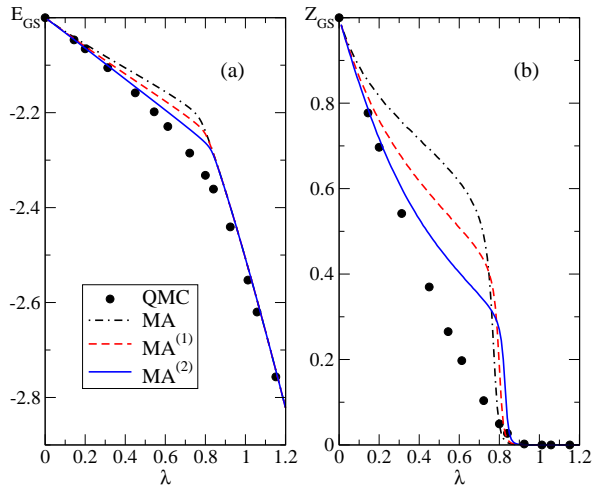}
\caption{(color online) (a) Ground state energy and (b) Ground state
  quasiparticle weight, as a function of the effective coupling
  $\lambda$, for $d=1, t=1, \Omega/t=0.1$. The QMC results are from
  Ref. \onlinecite{AlexPhD}.}
\label{fig2}
\end{figure}

The ground-state energies predicted by MA are quite accurate for a
large range of parameters. The accuracy is known to worsen as
$\Omega/t \rightarrow 0$, however even for $\Omega/t =0.1$ the MA
energies have less than 5\% 
relative error. On the other hand, the GS $qp$ weight for this low
$\Omega/t $ ratio is quite wrong for intermediary couplings $\lambda
\sim 1$, although it does become asymptotically exact, as
expected. The comparison with Quantum Monte Carlo (QMC) results is
shown in Fig. \ref{fig2}, 
where we also show the MA$^{(1)}$ and MA$^{(2)}$ predictions.

The accuracy is improved significantly for the higher MA
levels. Improvements are observed for all other sets of parameters
(not shown) analyzed in Ref.~\onlinecite{MA2}, however for higher
$\Omega/t$ ratios MA is much more accurate to begin with, so the
supplementary improvements due to MA$^{(1)}$ and MA$^{(2)}$ are
comparatively smaller. As discussed, this systematic improvement is
expected since all self-energy diagrams become more and  more
accurate. This is reflected in the sum rules for spectral weight,
which are also systematically improved (the link between diagrams and
sum rules is discussed at length in Ref.~\onlinecite{MA2}). While MA
satisfies the first 6 sum rules exactly, MA$^{(1)}$ satisfies the
first 8 sum rules exactly, MA$^{(2)}$ satisfies the
first 10 sum rules exactly, etc. Of course, the accuracy of all 
higher order sum rules is also systematically improved by the use of
more accurate expressions for the diagrams.

This systematic improvement can also be understood in variational terms. As
pointed out in Ref. \onlinecite{Bar}, MA is equivalent to a variational
approach where the eigenfunctions are built within a restricted
Hilbert space that allows phonons only at the
site where the electron is. In other words, the real-space counterpart
of the generalized Green's functions $F_n$, namely $F_n(i; j, j_1,
\dots, j_n; \omega) = \langle 0 | c_i \hat{G}(\omega) c_j^\dag
b^\dag_{j_1} \cdots  b^\dag_{j_n}|0\rangle\rightarrow
F^{(0)}_n(i;j;\omega)\prod_{k=1}^{n} 
\delta_{j, j_k} $ at the $MA^{(0)}$ level. This is equivalent to
asking that the single-polaron eigenstates have
non-zero overlap only with basis states of the general type $c_j^\dag
(b_j^\dag)^n|0\rangle $, $\forall j, n$. It is straightforward to
verify that with this restriction, the resulting equations for $F_n$
lead to the MA$^{(0)}$ self-energy.

This observation immediately  explains the absence of the
polaron+one-phonon continuum, since within this restricted Hilbert
space it is not 
allowed to have a phonon far from where the electron is. To compensate
for the missing continuum's weight and satisfy the sum rules, the GS
$qp$ weight is increased within MA, as seen in Fig. \ref{fig2}(b) and
many other examples discussed in Ref. \onlinecite{MA2}.

In this variational interpretation, MA$^{(1)}$ almost corresponds to
using a restricted Hilbert space enlarged by basis states of the form
$c_j^\dag(b_j^\dag)^nb^\dag_{j'}|0\rangle $, $\forall j,j', n$,
i.e. it also includes states where one phonon could exist anywhere
with respect to the electron. The equivalence is not exact, because
the resulting equation for $F_2$ is not the same in the two cases (all
other equations for all other $F_n$ with $n\ne 2$ are the same). More
precisely, one should think of $MA^{(1)}$ as a variational method also
accompanied by a change in the Hamiltonian if and only if acting on
electron+two-phonon states.  Similarly, MA$^{(2)}$ corresponds to
using a restricted Hilbert space spanned by basis states that allow
any number of phonons on the electron site plus up to two phonons
anywhere else in the system, accompanied by a change of the
Hamiltonian if and only if acting on electron+three-phonon states (now
the equation for $F_3$ is not quite the same as in MA$^{(2)}$),
etc. Note that one 
could also define improvements to MA based on the variational
equations for $F_2$ (instead of that resulting from MA$^{(1)}$), $F_3$
(instead of 
MA$^{(2)}$), etc. However, these equations are more involved than the
corresponding MA$^{(n)}$ type equations, making the solution of the
system of coupled equations more difficult.

\begin{figure}[t]
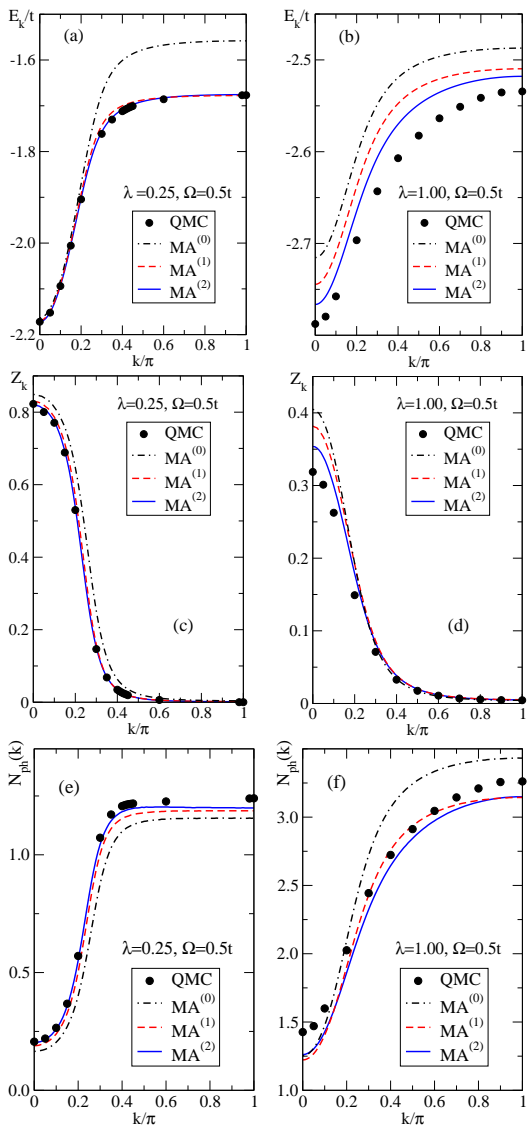

\includegraphics[width=0.80\columnwidth]{fig3a.eps}
\includegraphics[width=0.80\columnwidth]{fig3b.eps}
\includegraphics[width=0.80\columnwidth]{fig3c.eps}
\caption{(color online) (a) and (b) Polaron dispersion $E_k$; (c) and
  (d) $qp$ weight $Z_k$, and (e) and (f) average number of phonons
  $N_{\rm ph}(k)$ vs. $k$, in $d=1$, for $\Omega=0.5t$ and
  $\lambda=0.25$ in (a),(c)(e), respectively $\lambda=1.00$ in
  (b),(d),(f). The QMC results are from Ref. \onlinecite{AlexPhD}.}
\label{fig3}
\end{figure}

This systematic increase in the size of the variational Hilbert space
is another way to explain the gradual improvement of the GS energy. Also, it is now
clear that the polaron + one-phonon continuum should appear at level
MA$^{(1)}$ (see below). As a result, the GS $qp$ weight no longer
needs to account for it and it decreases, improving the agreement
with QMC results as shown in Fig. \ref{fig2}(b). We will return to this
issue when we investigate spectral weights. For the time being, we
note that in the adiabatic limit  $\Omega/t\rightarrow 0$, many
phonons can be created in 
the system at low energetic cost. In the intermediary region $\lambda
\sim 1$ where the polaron cloud is still relatively large, one expects
many of these phonons to be relatively far from the polaron and
therefore a high order $n$ would be required in order to accurately
describe them within this approach. As a result, it is expected that
the strongly adiabatic regime will not be quantitatively well
described for $\lambda \sim 1$ by the low-level MA approximations,
even though the qualitative behavior is correctly captured. Of
course, this limit can be investigated by other means, such as in
Ref. \onlinecite{barisic} and references therein.
However, for most of the parameter space, i.e.  any $\Omega/(dt) >
0.1$ or so and any coupling $\lambda$, the MA set of approximations give very
easy to evaluate yet remarkably accurate results.

\subsection{Polaron band}

We can also track how the lowest eigenstate of momentum ${\mb k}$, and
its various properties, change with various parameters. Results are
shown in Fig. \ref{fig3} for the energy $E_{\mb k}$, $qp$ weight
$Z_{\mb k}$ and average number of phonons $N_{\rm ph}({\mb k})$ for 1D
and two couplings, $\lambda = 0.25$ and $\lambda=1.00$. We found
similar improvements in 2D cases. Clearly,
MA$^{(2)}$ leads to an obvious improvement, even though
for this value of $\Omega/t=0.5$, MA itself is quite accurate
already. It should be noted that $\lambda \sim 1$ is where the MA
accuracy is generally expected to be at its worst.  In particular, for
the weak-coupling value $\lambda =0.25$, we see that MA overestimates
the distance to the continuum, i.e. $E_{\pi}-E_0> \Omega$. This should be
$\Omega$, but it is larger for MA because the polaron+one-phonon
continuum is not predicted at the correct energy.\cite{MA2} (For
$\lambda=1$, there is 
a second bound state between the one shown and the continuum,
therefore the bandwidth is much less than $\Omega$). For
MA$^{(1)}$ and MA$^{(2)}$ this problem is indeed fixed, and the
polaron dispersion width (at weak couplings) is $\Omega$. All other
quantities are also clearly more accurate. 

\subsection{Polaron+one-phonon continuum and higher energy states}

\begin{figure}[t]
\includegraphics[width=0.90\columnwidth]{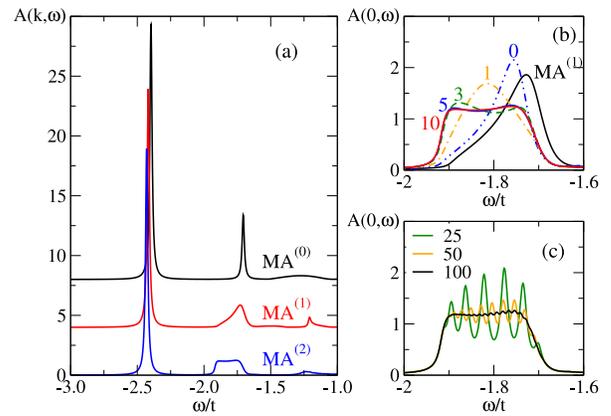}
\caption{(color online) (a) Spectral weight $A(k=0,\omega)$
  vs. $\omega$ in 1D for $t=1, 
  \Omega=0.5, \lambda = 0.6, \eta=0.01$, in MA, MA$^{(1)}$ and
  MA$^{(2)}$ (curves are shifted for clarity); (b) Polaron+one-phonon
  continuum convergence with cutoff 
  within MA$^{(2)}$. Results for cutoffs of 0, 1, 3, 5 and 10 are
  shown (the last two are almost identical). For comparison, the
  MA$^{(1)}$ continuum is also shown (black full line); (c) same as in
  (b), but for an inefficient computation scheme.}
\label{fig4}
\end{figure}

In order to understand the effects on higher-energy states, we study the spectral
  weight $A({\mb k}, \omega) = -{1\over 
  \pi} \mbox{Im} G({\mb k}, \omega)$. As is well known, this is finite
  only at energies $\omega$ where eigenstates of momentum
  $\mb{k}$ exist. For discrete (bound)
  states the spectral weight is a Lorentzian of width $\eta$ and
  height proportional to the $qp$ weight. In a continuum, the lifetime is
  determined by Im$\Sigma(\mb{k},\omega)$ and is independent
  of $\eta$ if $\eta$ is chosen small enough.

  In Fig. \ref{fig4}(a) we show results for the 1D spectral weight
  $A(k=0,\omega)$ vs. $\omega$ for a relatively weak coupling $\lambda
  = 0.6$. The MA spectral weight shows two discrete states at low
  energies, and a continuum starting for $\omega > -1.5 t$. Within
  MA$^{(1)}$, the second peak spreads into a continuum whose lower edge
  is at roughly $\Omega$ above the energy of the GS peak. In fact,
  since $\bar{g}_0(\omega)$ acquires an imaginary part when $-2dt \le
  \omega \le 2dt$, from Eq. (\ref{ma1}) it follows that the
  MA$^{(1)}$ continuum appears when $\tilde{\omega} \ge -2t
  \rightarrow \omega > E_{GS}^{(0)} + \Omega$, where $E_{GS}^{(0)}$ is
  the MA prediction for the GS energy. Since the MA$^{(1)}$ GS energy
  $E_{GS}^{(1)}<E_{GS}^{(0)}$, it follows that the gap is in fact slightly
  larger than $\Omega$.
  In MA$^{(2)}$ the weight is redistributed within this
  continuum, and its lower band-edge is at $\Omega$ above the
  ground-state energy, within numerical precision. 

The convergence of MA$^{(2)}$ with the cutoff value used to truncate
Eqs. (\ref{x1}) is shown in Fig.  \ref{fig4}(b). For a cutoff value of
0 we obtain the momentum-independent self-energy of Eq. (\ref{x2}), which
gives a continuum with a shape rather similar to that predicted by
MA$^{(1)}$. As the cutoff value is increased, weight shifts
towards the lower band-edge. Convergence is reached very quickly, with
little difference visible between results corresponding to a cutoff of
5 or 10 (these imply solving an inhomogeneous system of 11,
respectively 21 equations in Eqs. (\ref{x1})). Other solutions
of the coupled equations (briefly discussed in the previous section)
converge much more slowly. In Fig.  \ref{fig4}(c) we show results for
3 cutoffs for such an alternative scheme. Even for a cutoff as large
as 100, one can still see small oscillations,
very reminiscent of finite-size effects. The finding of an efficient
solution for the MA$^{(2)}$ self-energy 
is thus quite important.

\begin{figure}[t]
\includegraphics[width=0.80\columnwidth]{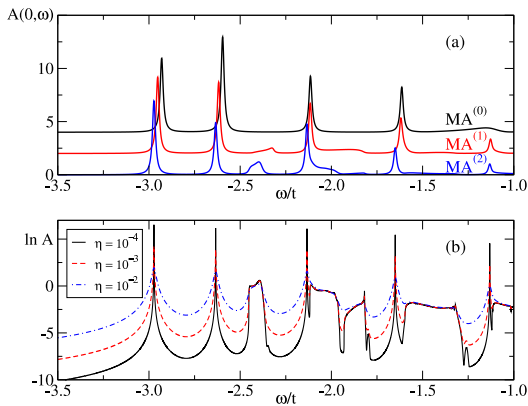}
\caption{(color online) (a) Spectral weight $A(0,\omega)$
  vs. $\omega$ in 1D for $t=1, 
  \Omega=0.5, \lambda = 1.2, \eta=0.01$, in MA, MA$^{(1)}$ and
  MA$^{(2)}$ (curves shifted for clarity); (b) $\ln A(0,\omega)$
  vs. $\omega$ for MA$^{(2)}$ and $\eta = 10^{-2}, 10^{-3},
  10^{-4}$. Other parameters are as in (a).}
\label{fig5}
\end{figure}

\begin{figure}[t]
\includegraphics[width=0.80\columnwidth]{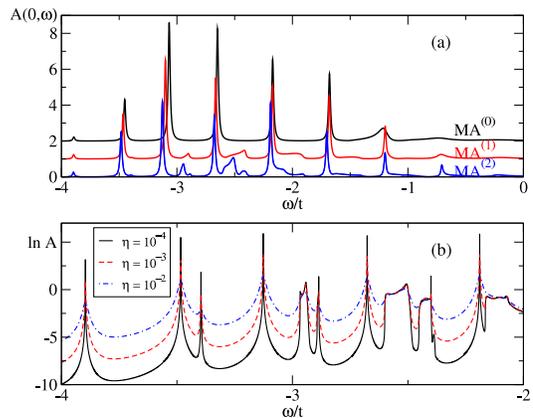}
\caption{(color online) (a) Spectral weight $A(0,\omega)$
  vs. $\omega$ in 1D for $t=1, 
  \Omega=0.5, \lambda = 1.8, \eta=0.01$, in MA, MA$^{(1)}$ and
  MA$^{(2)}$ (curves shifted for clarity); (b) $\ln A(0,\omega)$
  vs. $\omega$ for MA$^{(2)}$ and $\eta = 10^{-2}, 10^{-3},
  10^{-4}$. Other parameters are as in (a).}
\label{fig6}
\end{figure}

Similar behavior is observed for higher couplings, as seen in
Figs. \ref{fig5}(a) and \ref{fig6}(a) for intermediate, respectively
strong couplings $\lambda=1.2$ and 1.8.  In both cases there are now
two bound, discrete states below the continuum starting at
$E_{GS}+\Omega$, as expected.\cite{bonca:1999} The weight of the
polaron+one-phonon continuum decreases very fast, so that for
$\lambda=1.8$ it is barely 
visible just above the second peak. Fig. \ref{fig6}(b) shows it
clearly, on a logarithmic scale. Interestingly, it is not only the
height of this feature that is much smaller as $\lambda$ increases,
but its width as well. In 
fact, scaling vs. $\eta$ in Fig. \ref{fig6}(b) shows that this is more like
a Lorentzian, i.e. a single bound state, and not a finite-width
continuum as was the case for lower couplings (see
Fig. \ref{fig5}(b)). This is in fact reasonable, since at such large
couplings the lowest energy polaron state is basically dispersionless
(the effective mass is already considerable and the polaron is well
into the small-polaron regime). Since the width of the continuum is
due to the polaron dispersion (the phonon being dispersionless) it is
reasonable that as the polaron bandwidth decreases exponentially with
increasing coupling, so does the width of the polaron+one-phonon
continuum.

The higher energy features are also quite interesting. For the
intermediate coupling $\lambda=1.2$, at some distance above the
polaron+one-phonon continuum one can see the feature evolved from what
was the third discrete state in the MA approximation. For a large
$\eta$ value this looks like a continuum, however scaling with $\eta$
reveals a discrete state just below another continuum. In fact, the
spectrum is broken up into discrete states and continua separated by
gaps where no states are present. Of course, the detailed shape of the
spectral weight above the third bound state is likely to change as one
goes to MA$^{(3)}$ and higher orders, however it seems implausible
that these gaps should all close and a single continuum should form
above $E_{GS}+\Omega$, as is the case at very weak couplings (see
below). Instead, these results suggest that most weight is inside
bound states which are reminiscent of the Lang-Firsov ``comb'' of
discrete states separated by a frequency $\Omega$. Here the distance
between discrete states is generally less than $\Omega$ and there are
narrow continua in between them, which evolve from lower-energy
bound-states + one or more phonons. This is nicely illustrated by the
results in Fig. \ref{fig6}(b), which show 2 different features between
the third and the fourth bound states. The lower-energy one is a
continuum that starts roughly at $\Omega$ above the second bound
state, and has a finite width. Since the second bound state still has
some finite bandwidth at this coupling (see below), it seems
reasonable to interpret this feature as being the polaron in the
second-bound state plus one phonon somewhere far from it. The higher
energy feature is much narrower, but close inspection reveals that it
also is a continuum, with the lower edge starting at $E_{GS}+2\Omega$,
so its origin is obvious. The continua between higher bound states
become wider, in agreement with expectations if one assumes that
indeed they result from adding a distant phonon to a polaron in a
higher bound state, which has a larger bandwidth. There is also
overlap between different types of states, which also leads to
increased bandwidth.

\begin{figure}[t]
\includegraphics[width=0.8\columnwidth]{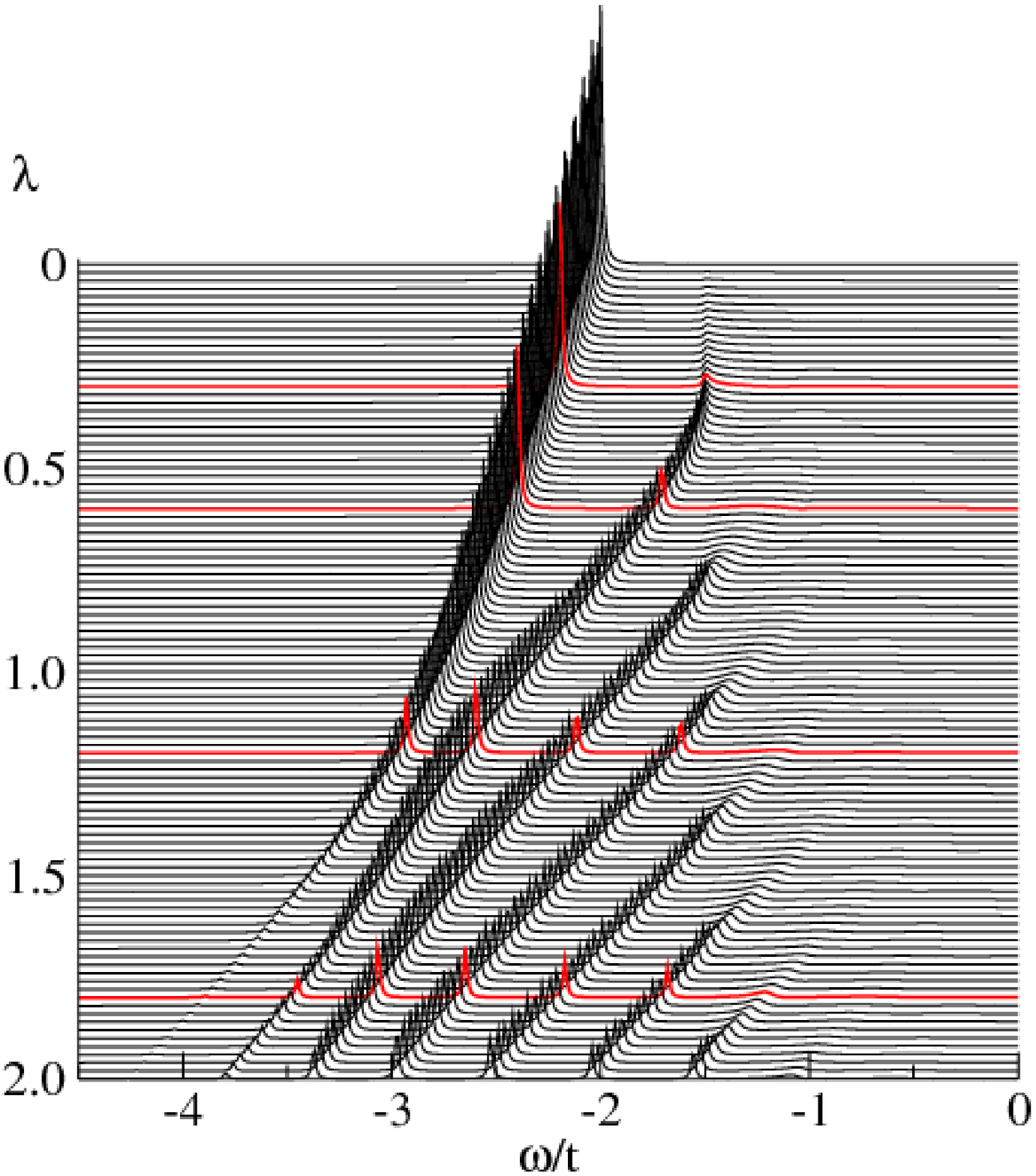}
\includegraphics[width=0.8\columnwidth]{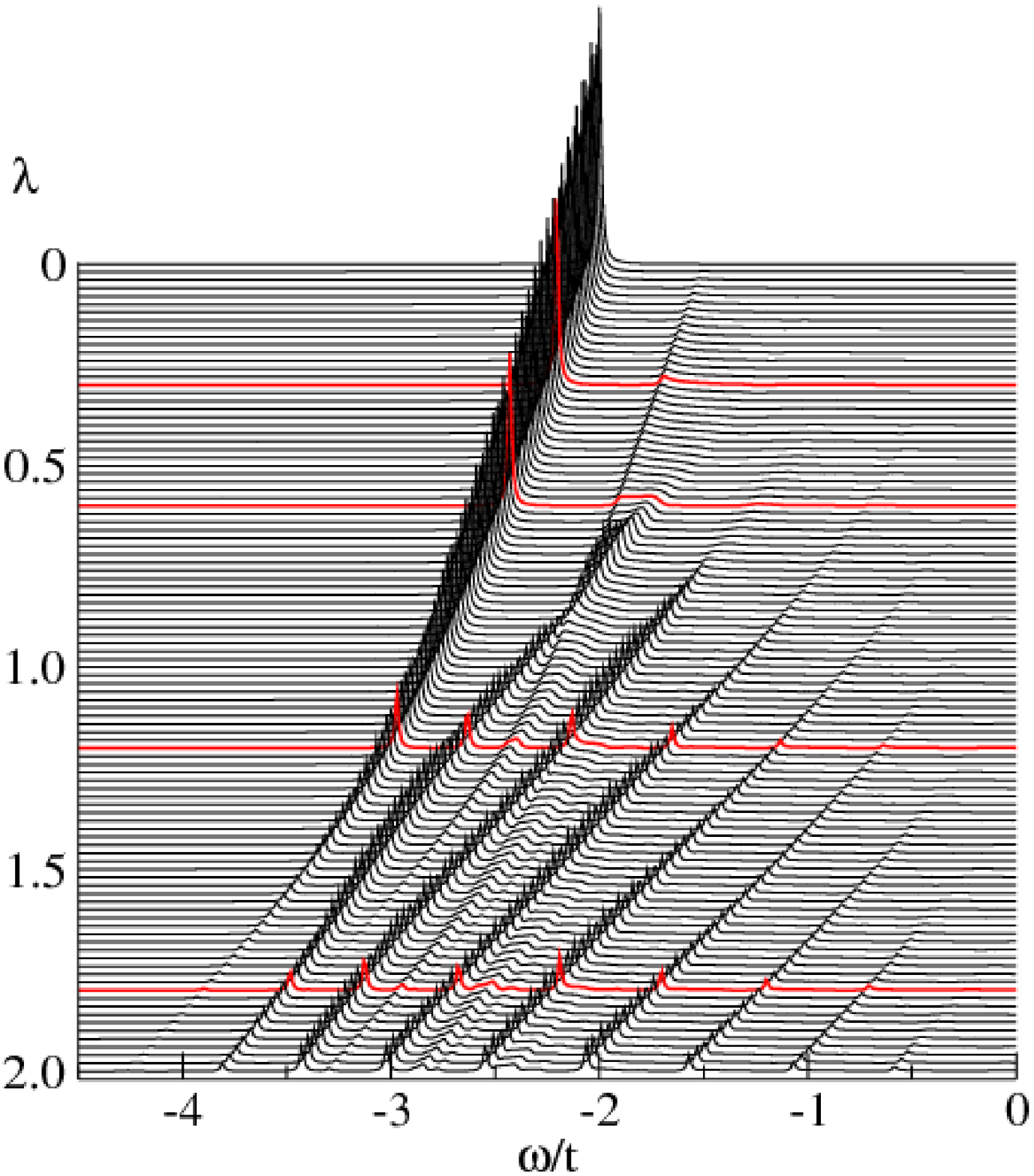}
\caption{(color online) (top) MA spectral weight $A(0,\omega)$
  vs. $\omega$ in 1D for $t=1, 
  \Omega=0.5,\eta=0.01$ and $\lambda$ varying from 0 to 2; (bottom) Same
  for MA$^{(2)}$. Curves corresponding to $\lambda=0.3, 0.6, 1.2$ and
  1.8 are highlighted.}
\label{fig7}
\end{figure}

This structure of the spectral weight explains why the MA (which
predicts only bound states at low energies, for medium to strong
coupling) still obeys sum rules with such good accuracy.\cite{MA2}
Most of the weight is indeed in the bound states, not in the narrow
continua that appear in between them. These results also show clearly
how the convergence towards the Lang-Firsov limit $g \gg t$ is
achieved: the width and weight of the continua shrinks to zero, and
one is left only with the equally spaced discrete states. This is
illustrated in Fig. \ref{fig7}, where $k=0$ MA spectral weights are
contrasted with MA$^{(2)}$ spectral weights for different $\lambda$
values. The largest difference is observed at small couplings, where
MA predicts the (wrong) continuum pinned at $-2dt + \Omega = -1.5$ for
these values, whereas MA$^{(2)}$ clearly shows a continuum starting at
$\Omega$ above the ground-state for as long as it is still 
visible. For $\lambda > 0.7$ or so, a second bound state splits from
this continuum and comes quite close to the GS peak before moving away
so that it asymptotically goes to $E_{GS}+\Omega$. There are clear
similarities between the two plots, with most of the weight
concentrated in the bound states that are roughly $\Omega$ apart, but
MA overestimates their weights in order to compensate for the missing
narrow continua that appear in between these discrete states. As
stated, the precise shape and weight of the higher continua is very
likely to change as one goes to a higher level MA approximation, however
we expect the general picture to remain essentially the same.

We are unable to check  quantitatively the  accuracy of the higher energy
spectral weight against detailed numerical predictions, beyond the
sort of comparisons shown for the polaron dispersion in
Fig. \ref{fig3}. The reason is that most of the numerical
work is focused on computing low-energy properties (a detailed
overview of such work is provided in Ref. \onlinecite{MA2}, or, for
example, in Ref. \onlinecite{numer_gs}). The
much fewer high-energy results, such as those based on a
variational treatment in Ref. \onlinecite{cataud} and a novel QMC /
exact diagonalization approach in Ref. \onlinecite{hohen2} use a
rather large $\eta$ and are already in reasonable agreement with MA,
as discussed in Ref. \onlinecite{MA2}. Cluster perturbation theory
(CPT) results such as shown in Ref. \onlinecite{hohen1}, while also an
approximation, are in good qualitative and quantitative agreement with
ours. It would be very interesting to be able to compare our results
against detailed high-accuracy, high-energy numerical predictions.

\begin{figure}[t]
\includegraphics[width=0.95\columnwidth]{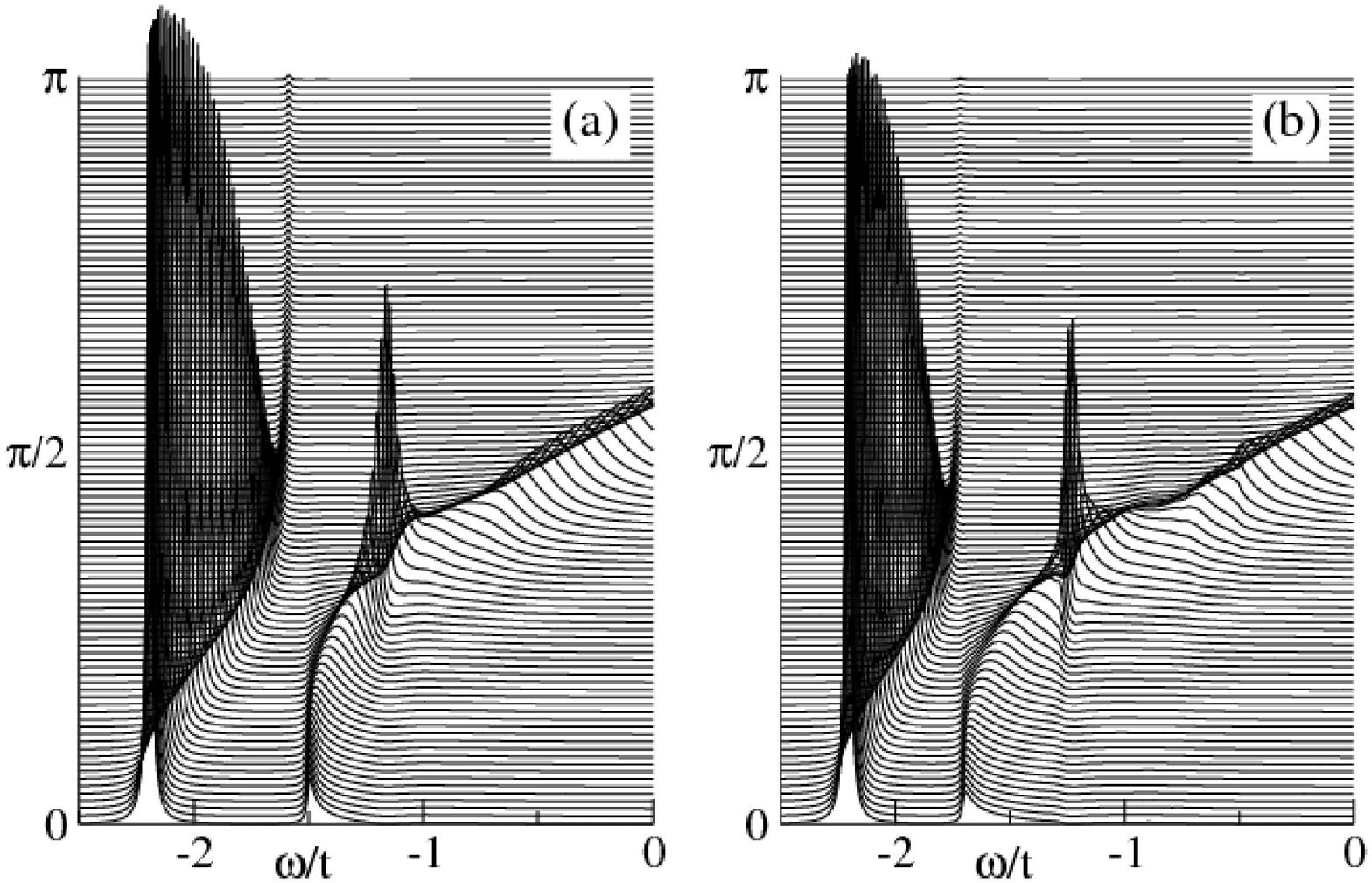}
\includegraphics[width=0.95\columnwidth]{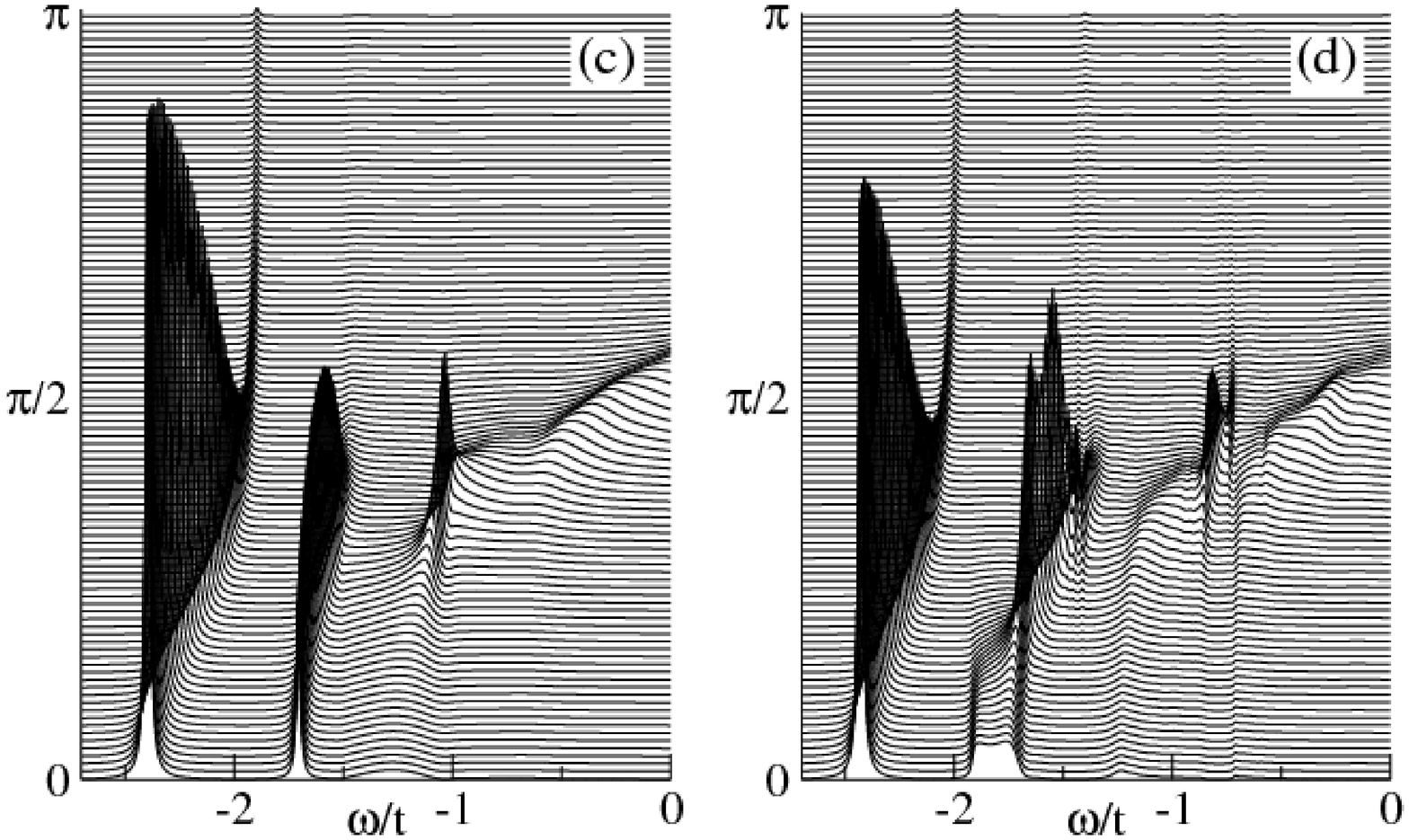}
\caption{$A(k,\omega)$
  vs. $k$ and $\omega$ in 1D for $t=1, 
  \Omega=0.5,\eta=0.01$ and $\lambda=0.3$ in (a), (b) and  $\lambda=0.6$
 in (c), (d). Results for MA are
  shown in (a), (c), while MA$^{(2)}$ is shown in (b), (d). }
\label{fig8}
\end{figure}

\begin{figure}[t]
\includegraphics[width=0.95\columnwidth]{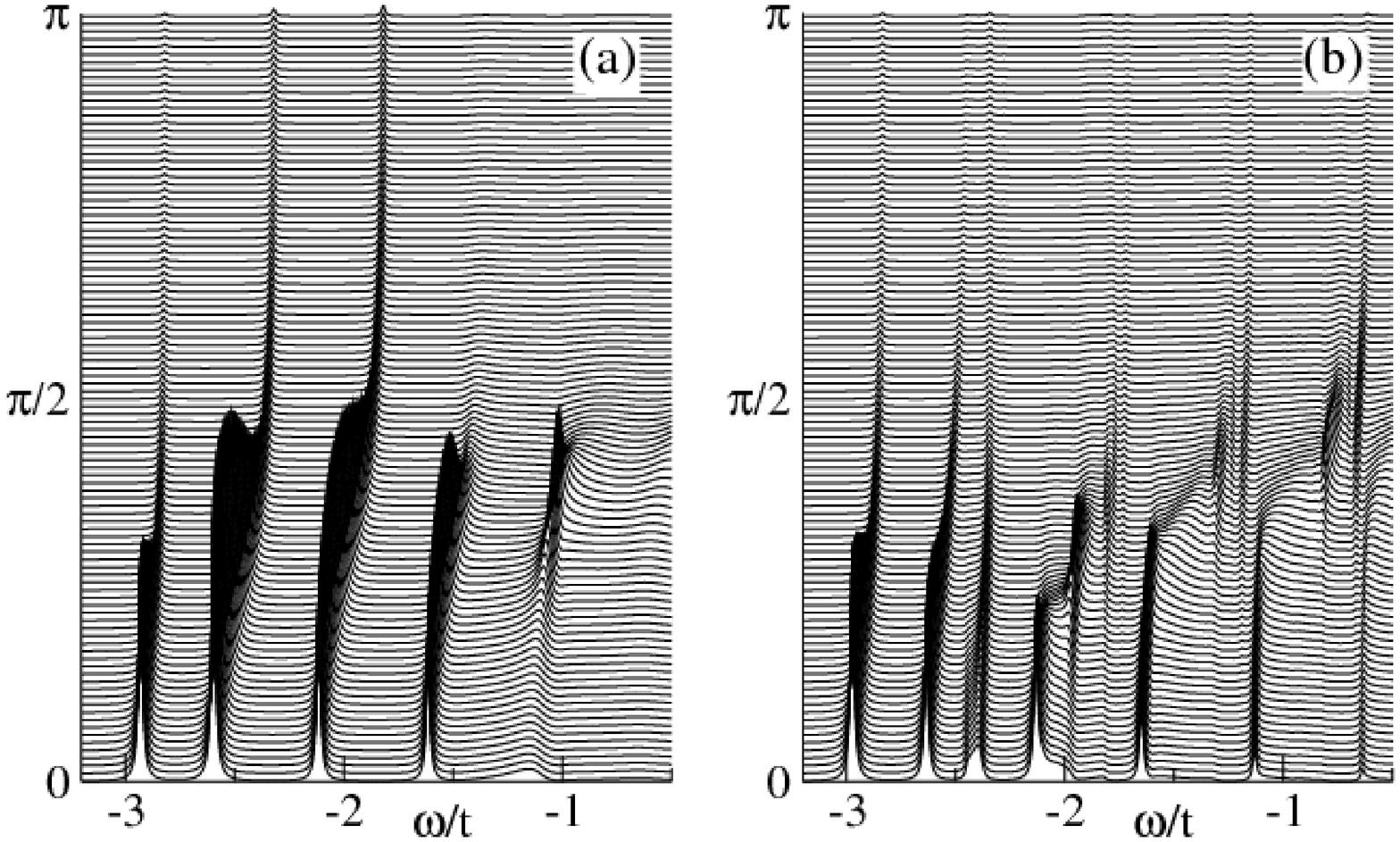}
\includegraphics[width=0.95\columnwidth]{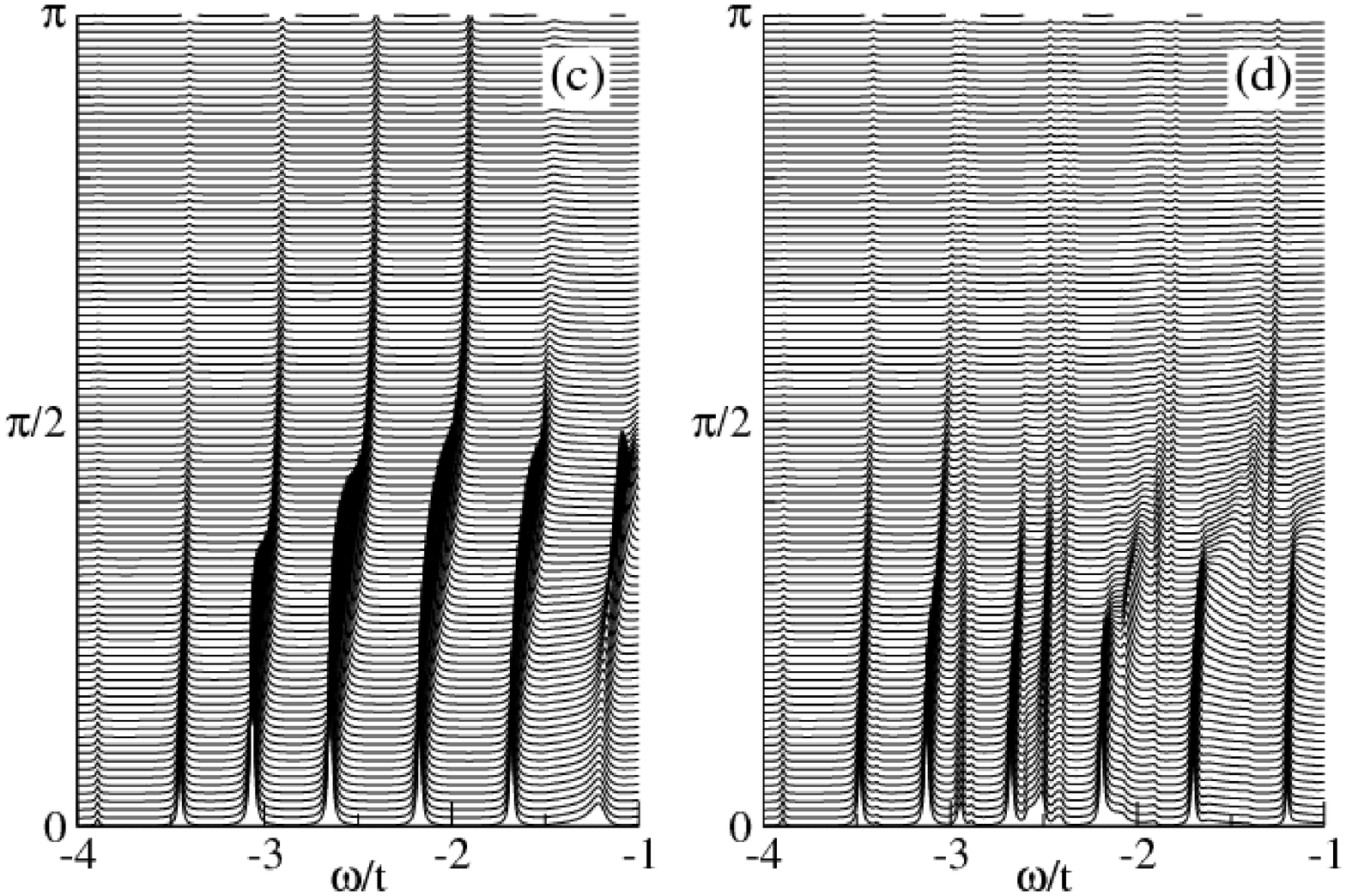}
\caption{$A(k,\omega)$
  vs. $k$ and $\omega$ in 1D for $t=1, 
  \Omega=0.5,\eta=0.01$ and $\lambda=1.2$ in (a), (b) and  $\lambda=1.8$
 in (c), (d). Results for MA are
  shown in (a), (c), while MA$^{(2)}$ is shown in (b), (d). }
\label{fig9}
\end{figure}

Finally, we contrast the difference between MA and MA$^{(2)}$ spectral
weights for different momenta and energies. Typical results are shown
in Figs. \ref{fig8} and \ref{fig9}, for weak, intermediate and strong
couplings $\lambda=0.3, 0.6, 1.2$ and 1.8, respectively. For the weak
coupling, as already discussed, the most obvious difference is that
the polaron bandwidth is decreased to its correct value of $\Omega$ in
MA$^{(2)}$. The $qp$ weights and all other features are very
similar. Based on the MA$^{(2)}$ results, it is now clear that the
strong resonance seen in the electron+phonon continuum occurs at an
energy of $2\Omega$ above the ground-state energy. One expects that
here is where the second bound state will arise from. For
$\lambda=0.6$, Figs. \ref{fig8}(c) and (d) show a bigger
contrast. Here, MA already predicts a second bound state that has
evolved in between the polaron band and the higher-energy
continuum. In contrast, MA$^{(2)}$ shows that there is no second-bound
state yet, however the electron+one-phonon continuum is split off the
higher energy continuum, which also starts to fractionalize at higher
energies (roughly multiples of $\Omega$). Within MA$^{(2)}$, a true
second bound-state is observed for the higher couplings shown in
Fig. \ref{fig9}. In contrast to MA, which shows several bound states
which disperse as $k$ increases, MA$^{(2)}$ also clearly shows
continua in between these discrete states. These account for some of
the spectral weight that was in the MA peaks. These results again
suggest a very fractionalized spectrum at intermediate and strong
couplings. Instead of the polaron band, a second-bound state and a
rather featureless continuum at higher energies, we instead find many
sets of discrete states interspersed with continua. As
$\lambda\rightarrow \infty$, the relative weight in these continua
decreases and the spectral weights evolve continuously towards the
Lang-Firsov set of discrete states with energies $-g^2/\Omega +
n\Omega$.

Qualitatively similar results are observed in higher
dimensions. Rather ironically, the most time-consuming part in the
MA$^{(2)}$ calculation for higher dimensions is finding the spatial
dependence of the non-interacting Green's functions $G_0(i,\Omega)$
which are needed to generate Eqs. (\ref{x1}), and not the  solving of the
system of equations. The reason is that for nearest-neighbor hopping in
higher dimensions, the evaluation of these propagators must be done by
numerically. Of course, one could choose simpler forms of
the dispersion $\epsilon_{\mb k}$ for which analytical results are
possible. However, as we show now, in higher dimensions the
improvements in going to MA$^{(1)}$ and MA$^{(2)}$ are quantitatively
smaller, because the relative weight in the continua is reduced compared to the
1D case. This is in agreement with our general observation that MA
itself becomes more accurate with increased dimensionality.\cite{MA2} 

\begin{figure}[t]
\includegraphics[width=0.80\columnwidth]{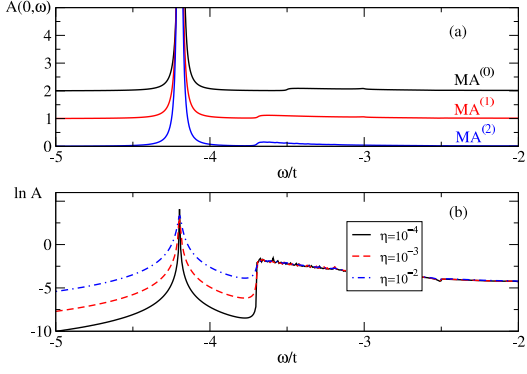}
\caption{(color online) (a)  $A(0,\omega)$
  vs. $\omega$ in 2D for $t=1, 
  \Omega=0.5, \lambda = 0.3, \eta=0.01$, in MA, MA$^{(1)}$ and
  MA$^{(2)}$ (curves shifted for clarity); (b) $\ln A(0,\omega)$
  vs. $\omega$ for MA$^{(2)}$ and $\eta = 10^{-2}, 10^{-3},
  10^{-4}$. Other parameters are as in (a).}
\label{fig10}
\end{figure}

\begin{figure}[b]
\includegraphics[width=0.80\columnwidth]{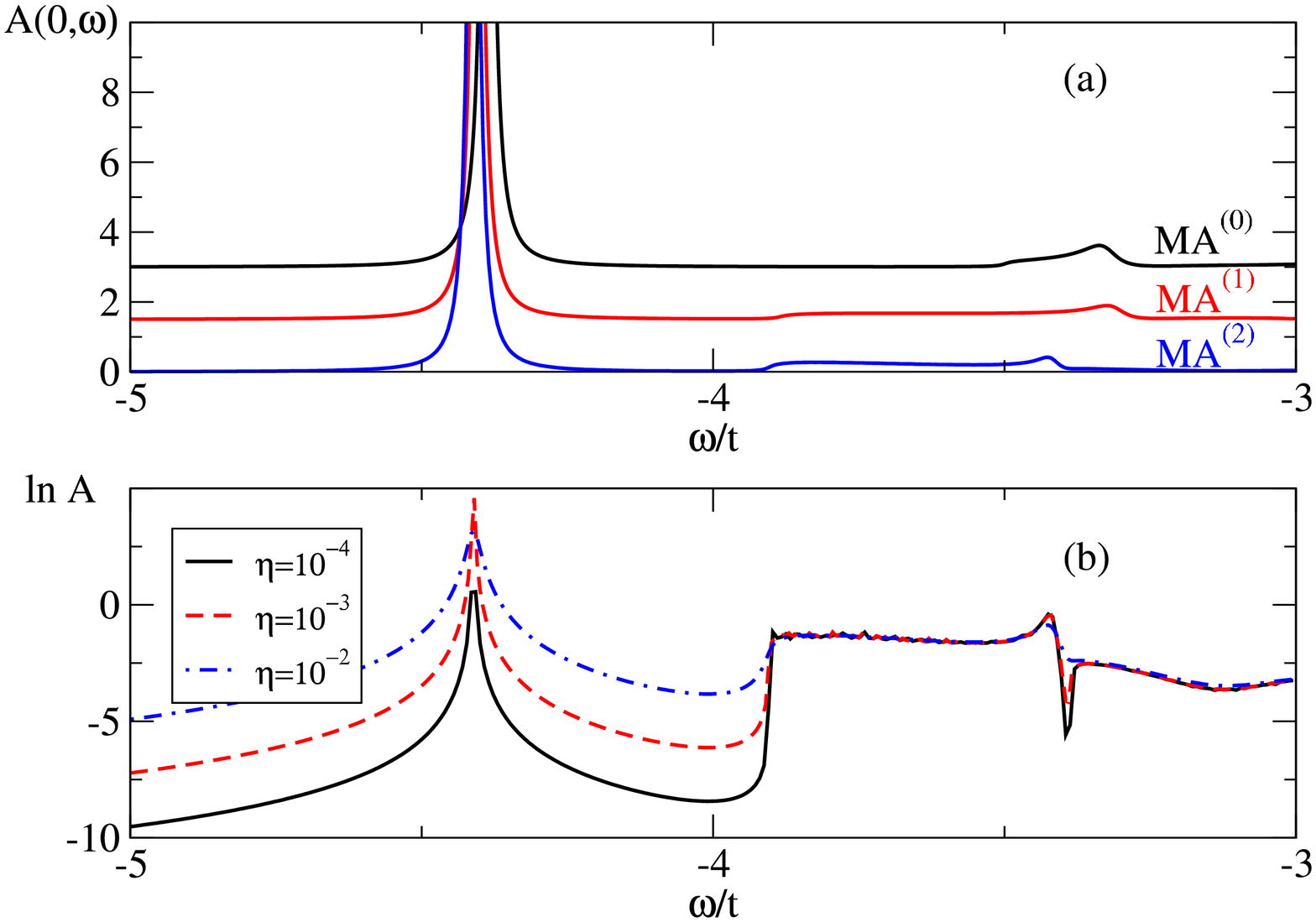}
\caption{(color online) (a) $A(0,\omega)$
  vs. $\omega$ in 2D for $t=1, 
  \Omega=0.5, \lambda = 0.6, \eta=0.01$, in MA, MA$^{(1)}$ and
  MA$^{(2)}$ (curves shifted for clarity); (b) $\ln A(0,\omega)$
  vs. $\omega$ for MA$^{(2)}$ and $\eta = 10^{-2}, 10^{-3},
  10^{-4}$. Other parameters are as in (a).}
\label{fig10a}
\end{figure}

\begin{figure}[t]
\includegraphics[width=0.80\columnwidth]{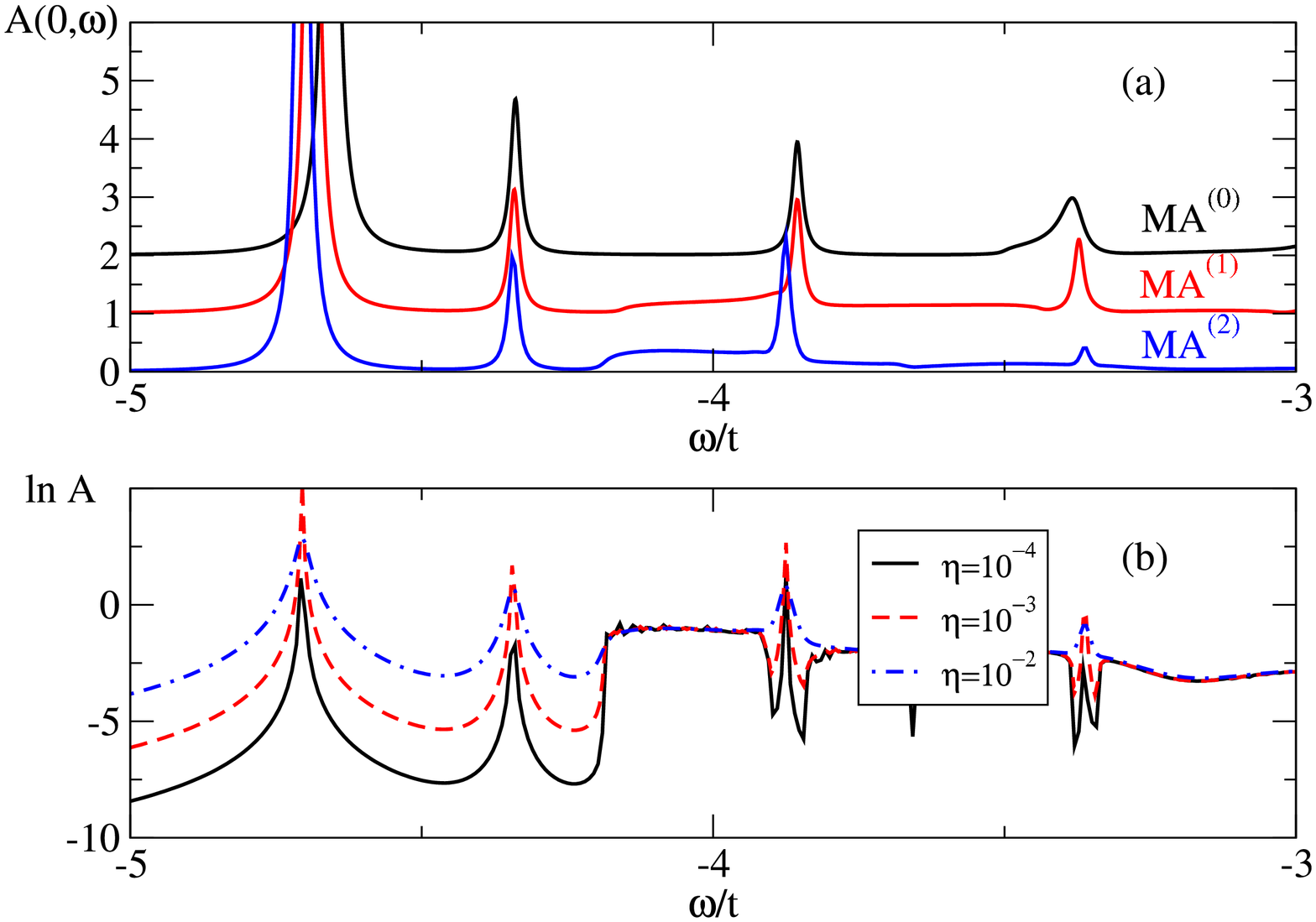}
\caption{(color online) (a)  $A(0,\omega)$
  vs. $\omega$ in 2D for $t=1, 
  \Omega=0.5, \lambda = 0.95, \eta=0.01$, in MA, MA$^{(1)}$ and
  MA$^{(2)}$ (curves shifted for clarity); (b) $\ln A(0,\omega)$
  vs. $\omega$ for MA$^{(2)}$ and $\eta = 10^{-2}, 10^{-3},
  10^{-4}$. Other parameters are as in (a).}
\label{fig11}
\end{figure}

\begin{figure}[b]
\includegraphics[width=0.80\columnwidth]{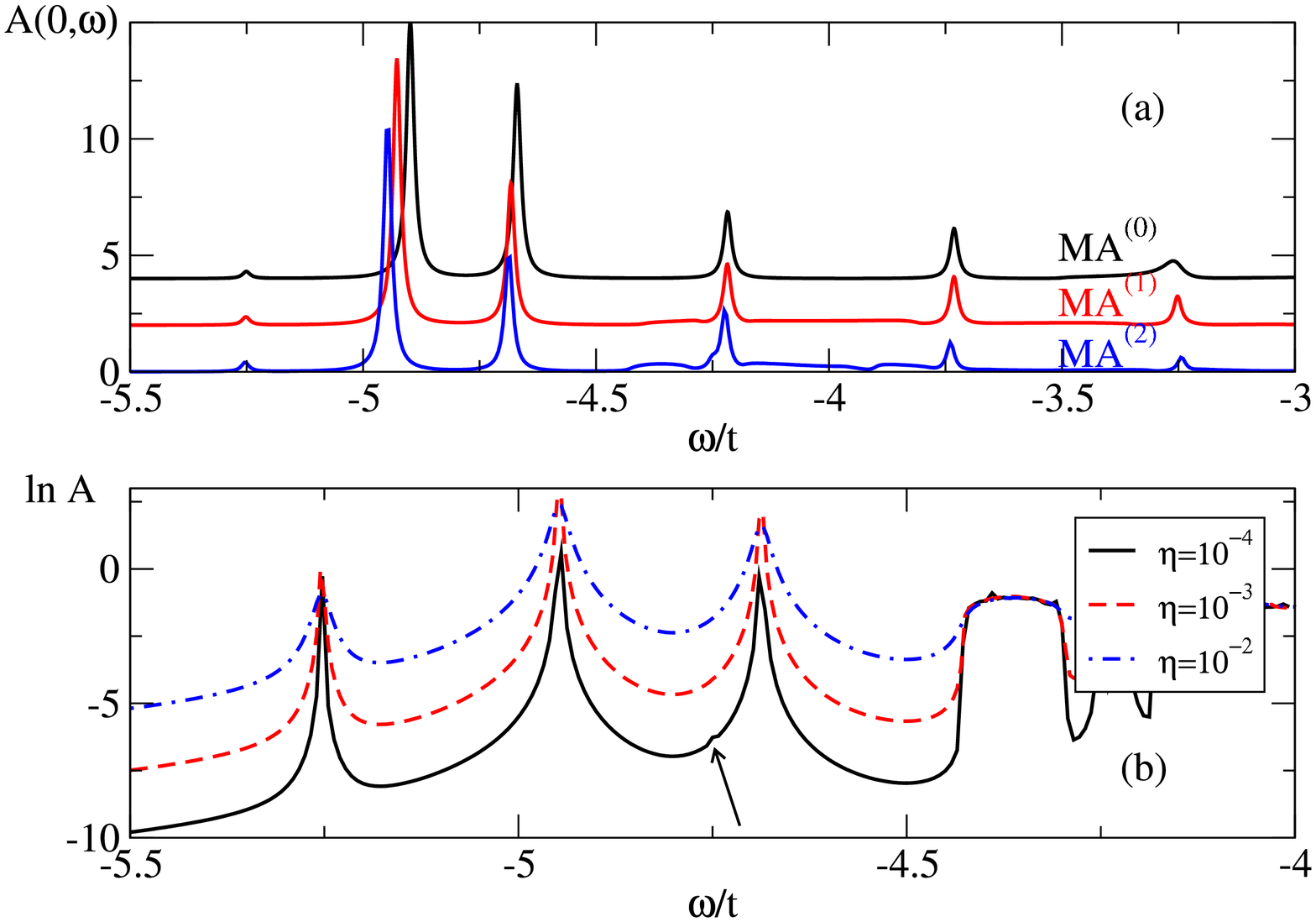}
\caption{(color online) (a) $A(0,\omega)$
  vs. $\omega$ in 2D for $t=1, 
  \Omega=0.5, \lambda = 1.2, \eta=0.01$, in MA, MA$^{(1)}$ and
  MA$^{(2)}$ (curves shifted for clarity); (b) $\ln A(0,\omega)$
  vs. $\omega$ for MA$^{(2)}$ and $\eta = 10^{-2}, 10^{-3},
  10^{-4}$. Other parameters are as in (a).}
\label{fig12}
\end{figure}

In Figs. \ref{fig10}-\ref{fig12} we show the 2D spectral weight
$A({\mb k}=0, \omega)$ vs. $\omega$, for effective couplings
$\lambda=0.3, 0.6, 0.95$ and 1.2, both on linear and logarithmic
scales. Qualitatively, everything is similar to the behavior seen in
the 1D case. Quantitatively, we find that the continuum that appears
at $\Omega$ above the ground-state is broader, but of lower height. In
fact, its height is so small that it is invisible on curves
like those in Fig. \ref{fig7}, which is the reason why we do not show
such curves here. The overall weight in this continuum also decrease
much faster with increasing $\lambda$. In 1D, for $\lambda=1.2$ the
first continuum is still clearly visible, see Fig. \ref{fig5}. For
$\lambda=1.8$ it becomes harder to see  on the linear scale, but it
is clearly seen on the logarithmic scale. By contrast, in 2D, for
$\lambda=1.2$ the continuum is no longer visible on the linear scale,
and even in the logarithmic scale it only barely starts to be visible
for $\eta = 10^{-4}$ (small shoulder marked by arrow). The spectral
weight for a Lorentzian contribution $Z/(\omega +i \eta)$ is
$Z\eta/[\pi(\omega^2 + \eta^2)]$, so the maximum height of the
peak, at resonance, is $Z/\pi\eta$. This is visible only if it is
larger than the background, which of course depends on how close is
the next spectral feature. However, for a very small $Z$, $\eta$ has to
be really small before the peak is seen.

As in the 1D case, we also observe the fractionalization of the
spectrum for moderate and large effective couplings, with a succession
of discrete peaks and continua at higher energies. We expect 
similar behavior to be observed in 3D as well. Clearly,  the changes
in going from MA to MA$^{(1)}$ and  MA$^{(2)}$ 
are quantitatively much smaller in 2D than in 1D, because the continua
have so little weight. We expect the trend to continue in going to 3D,
meaning that in 3D, the difference between MA$^{(2)}$ and MA should be
quantitatively even less. Indeed, all the comparisons of MA results
with available 3D numerics, shown in Ref. \onlinecite{MA2}, are already in
excellent agreement, even for intermediary couplings $\lambda \sim
1$. As a result, we find it unnecessary to present 3D results,
although they can be obtained very straightforwardly.

\section{Conclusions}

In summary, we presented a way to systematically improve the MA
approximation, by systematically improving the accuracy of
self-energy diagrams, but in such a way that they can still all be
efficiently summed. 

This allows us to rather easily fix various known failings of the MA
approximation, such as  the absence of the polaron+one-phonon continuum at the
correct energy, and its momentum-independent self-energy. It also
allows us to understand in more detail the effects of the
Holstein-type electron-phonon coupling on the polaron spectrum, both
at low and high energies. While agreement with exact numerical results
is improved, unfortunately there are not many such results for higher
energy states, so detailed comparisons are not possible there. However, the
hierarchy of MA$^{(n)}$ approximations is clearly providing a simple
 way towards obtaining quantitatively more and
more accurate results for the Green's function of the Holstein
polaron, in any dimension and for any free electron dispersion. 

The next direction of obvious interest is to study to what extent
this work can be 
extended to other models, for example models where the electron-phonon
coupling depends on the phonon  momentum ({\em e.g.} the
breathing-mode\cite{Slezak} or Fr\"olich\cite{Fro}
Hamiltonians). For such models there are 
very few reliable high-energy results available. A simple
approximations such as MA could easily investigate the whole parameter
space and identify interesting regimes. Such work is currently in progress.

{\bf Acknowledgments:} We thank O. S. Bari\v{s}i\'c for useful
comments and  George A. Sawatzky for many discussions.  This work was supported
by the A. P. Sloan Foundation, CIfAR Nanoelectronics, NSERC and CFI.

\appendix

\section{Continued fractions}
\label{ap1}

Consider the set of recurrence relations $f_{-1}=1$ and for $n\ge 0$,
$$ f_n = \alpha_n f_{n-1} + \beta_n f_{n+1},
$$ where $f_n \rightarrow 0$ as $n\rightarrow \infty$ (this is
justified for all cases of interest to us since $f_n$ is related to a
generalized Green's function with $n$ phonons in the initial state and
none in the final state. If $n$ is large enough, the amplitude of
probability to evolve between such states must vanish).

Truncating the system at a large enough $N$ so that $f_{N+1} \approx
0$, we solve backwards to find $f_N = \alpha_N f_{N-1}$, $f_{N-1} =
\alpha_{N-1} f_{N-2}/(1- \beta_{N-1}\alpha_{N})$, .... In general,
$$ f_n =\cfrac{\alpha_{n} f_{n-1}}{1-\cfrac{\alpha_{n+1}\beta_n}{1-
\cfrac{\alpha_{n+2}\beta_{n+1}}{1-\dots}}} = A_n f_{n-1}.
$$ If we allow $N\rightarrow \infty$, then the solution becomes exact
and $A_n= \alpha_n/[1- \beta_n A_{n+1}]$ are infinite continued
fractions.

\section{MA$^{(1)}$ details}
\label{ap2}

First, we use Eqs. (\ref{e21}) to find how ${1\over N} \sum_{{\mb
  q}_2}^{} f_2^{(1)}({\mb q_1}, {\mb q_2})$ depends on
  $f_1^{(1)}(\mb{q}_1)$. This is then used in Eq. (\ref{e11}) to solve
  for $f_1^{(1)}(\mb{q}_1)$, and $\Sigma$ follows from Eq. (\ref{s1}).

We introduce the partial momentum averages:
\begin{equation}
\EqLabel{a1} \bar{f}_n({\mb q}_1) = {1\over N^{n-1}}\sum_{\mb{q}_2,
\ldots,\mb{q}_{n}}^{} f_n^{(1)}(\mb{q}_1, \ldots, \mb{q}_n),
\end{equation}
(we use the convention that $\bar{f}_1({\mb q}_1) = f_1^{(1)}({\mb
q}_1)$) and the full momentum averages:
\begin{equation}
\EqLabel{a2} {\cal F}_n= {1\over N^n} \sum_{\mb{q}_1,
\ldots,\mb{q}_{n}}^{} f_n^{(1)}(\mb{q}_1, \ldots, \mb{q}_n) = {1\over
N} \sum_{{\mb q}_1}^{}\bar{f}_n({\mb q}_1)
\end{equation}
so that $\Sigma_{MA^{(1)}} ({\bf k}, \omega) = {\cal F}_1$.

First, we average each of Eqs. (\ref{e21}) over all the corresponding
momenta, to find that for $n\ge 2$, ${\cal F}_n =
\bar{g}_0(\omega-n\Omega)[ n g^2 {\cal F}_{n-1} + {\cal
F}_{n+1}]$. This has the continued-fraction solution (see Appendix
\ref{ap1}):
\begin{equation}
\EqLabel{f1} {\cal F}_2= g^2 A_2(\omega){\cal F}_1
\end{equation}
where $A_2(\omega)$ is a continued fraction defined in Eq. (\ref{cf})
and ${\cal F}_1$ is still unknown.

We now average each of Eqs. (\ref{e21}) over all the momenta except
${\mb q}_1$, to find that for all $n\ge 2$, $ \bar{f}_n({\mb q}_1) =
\bar{g}_0(\omega-n\Omega)[g^2 {\cal F}_{n-1} + (n-1)g^2
\bar{f}_{n-1}({\mb q}_1) + \bar{f}_{n+1}({\mb q}_1)].  $ This can also
be solved by rewriting it in terms of $\delta \bar{f}_n({\mb q}_1) =
\bar{f}_n({\mb q}_1)- {\cal F}_n$. Using the recurrence relation for
${\cal F}_n$, we find that:
$$ \delta \bar{f}_n({\mb q}_1) = \bar{g}_0(\omega-n\Omega)[ (n-1)g^2
\delta \bar{f}_{n-1}({\mb q}_1) + \delta \bar{f}_{n+1}({\mb q}_1)].
$$ This also has solutions in terms of continued fractions, in
particular $\delta \bar{f}_2({\mb q}_1) = g^2 A_1(\omega- \Omega)
\delta \bar{f}_1({\mb q}_1)$. Using Eq. (\ref{f1}) we then find $
\bar{f}_2({\mb q}_1) = g^2 A_1(\omega- \Omega) {f}_1({\mb q}_1) +
g^2\left[A_2(\omega) - A_1(\omega-\Omega)\right]{\cal F}_1$. This
expression is now inserted in Eq. (\ref{e11}) to give an equation with
only $ {f}_1^{(1)}({\mb q}_1)$ and ${\cal F}_1 = {1\over N} \sum_{{\mb
q}_1}^{} {f}^{(1)}_1({\mb q}_1)$ as unknowns, which can be re-arranged
as:
\begin{eqnarray}
\nonumber \left[\left(G_0({\mb k}-{\mb q}_1,
  \omega-\Omega)\right)^{-1} - g^2 A_1(\omega - \Omega)\right]
  {f}^{(1)}_1({\mb q}_1) & &\\ \nonumber = g^2 + g^2\left[A_2(\omega)
  - A_1(\omega-\Omega)\right]{\cal F}_1. & &
\end{eqnarray}
But $[G_0({\mb k}, \omega)]^{-1} - a(\omega) = \omega - \epsilon_{\bf
  k} + i \eta - a(\omega) = [G_0({\mb k}, \omega- a(\omega))]^{-1}$
  and therefore we find:
$$ {f}^{(1)}_1({\mb q}_1) = G_0({\mb k}-{\mb q}_1, \omega-\Omega- g^2
A_1(\omega - \Omega))
$$
$$ \times \left[g^2 + g^2\left[A_2(\omega) -
A_1(\omega-\Omega)\right]{\cal F}_1\right].
$$ After momentum averaging over ${\mb q}_1$ on both sides, we find
${\cal F}_1= \Sigma$ [see Eq. (\ref{ma1})]. Based on the knowledge of
${f}^{(1)}_1({\mb q}_1)$ and the various partial and total averages,
one can now compute ${f}^{(1)}_2({\mb q}_1,{\mb q}_2)$, etc.

\section{MA$^{(2)}$ details}
\label{ap3}

As for MA$^{(1)}$, the strategy is to solve Eqs. (\ref{e32}) to find
how ${1\over N} \sum_{{\mb q}_3}^{} f_3^{(2)}( {\mb q_1}, {\mb q_2},
{\mb q_3})$ depends on $f_2^{(2)}( \mb{q}_1, {\mb q_2})$. Using this
in Eq. (\ref{e22}) allows us to solve it in conjunction with
Eq. (\ref{e31}) to find $f_1^{(2)}(\mb{q}_1)$ and therefore $\Sigma$.

We introduce various partial momentum averages:
$$ \bar{\bar f}_n({\mb q}_1, {\mb q}_2) = {1\over
N^{n-2}}\sum_{\mb{q}_3, \ldots,\mb{q}_{n}}^{}
f_n^{(2)}(\mb{q}_1,\mb{q}_2, \ldots, \mb{q}_n),
$$
$$ \bar{f}_n({\mb q}_1) ={1\over N^{n-1}}\sum_{\mb{q}_2,
\ldots,\mb{q}_{n}}^{} f_n^{(2)}(\mb{q}_1, \ldots, \mb{q}_n).
$$ where we take $\bar{\bar f}_1({\mb q}_1)=\bar{f}_1({\mb q}_1) =
f_1^{(2)}({\mb q}_1)$,$ \bar{\bar f}_2({\mb q}_1, {\mb q}_2)=
f_2^{(2)}(\mb{q}_1,\mb{q}_2) $. We also define the full momentum
averages:
\begin{equation}
\EqLabel{b2} {\cal F}_n= {1\over N} \sum_{\mb{q}_1,
\ldots,\mb{q}_{n}}^{} f_n^{(2)}(\mb{q}_1, \ldots, \mb{q}_n) = {1\over
N} \sum_{{\mb q}_1}^{}\bar{f}_n({\mb q}_1)
\end{equation}
The functions $\bar{\bar f}_n, \bar{f}_n$ and ${\cal F}_n$ should also
carry the upper label $(2)$ since their values are different than
those obtained within MA$^{(1)}$, however we do not write it
explicitly to simplify the notation somewhat.

The solutions for ${\cal F}_n$ and $\bar{f}_n$ proceed just as in the
MA$^{(1)}$ case, with one difference.  Averaging all equations
(\ref{e32}) over all corresponding momenta, we find that now only for
$n\ge 3$, ${\cal F}_n = \bar{g}_0(\omega-n\Omega)[ n g^2 {\cal
F}_{n-1} + {\cal F}_{n+1}]$, therefore this recurrence relation now
ends with:
\begin{equation}
\EqLabel{f3} {\cal F}_3= g^2 A_3(\omega){\cal F}_2.
\end{equation}
Similarly, after using the same approach as for MA$^{(1)}$, one finds
that $\delta \bar{f}_3({\mb q}_1) = \bar{f}_3({\mb q}_1) - {\cal F}_3
= g^2 A_2(\omega- \Omega) \delta \bar{f}_2({\mb q}_1)$ so that $
\bar{f}_3({\mb q}_1) = g^2 A_2(\omega- \Omega) \bar{f}_2({\mb q}_1) +
g^2\left[A_3(\omega) - A_2(\omega-\Omega)\right]{\cal F}_2$. Of
course, the functions ${\cal F}_2, \bar{f}_2({\mb q}_1)$ are still
unknown. Finally, we can proceed to calculate $\bar{\bar f}_3({\mb
q}_1, {\mb q}_2)$ which is needed in Eq. (\ref{e22}).

We momentum average Eqs. (\ref{e32}) over all momenta except ${\mb
q}_1, {\mb q}_2$, to find
\begin{eqnarray}
\nonumber & &\bar{\bar f}_n({\mb q}_1, {\mb q}_2) =
\bar{g}_0(\omega-n\Omega) \left[g^2 \left(\bar{f}_{n-1}({\mb q_1}) +
\bar{f}_{n-1}({\mb q_2})\right) \right.\\ \nonumber &
&\left. +(n-2)g^2 \bar{\bar f}_{n-1}({\mb q}_1, {\mb q}_2) + \bar{\bar
f}_{n+1}({\mb q}_1, {\mb q}_2) \right].
\end{eqnarray}
Let $\delta \bar{\bar f}_n({\mb q}_1, {\mb q}_2) = \bar{\bar f}_n({\mb
q}_1, {\mb q}_2) - \bar{f}_{n}({\mb q_1}) - \bar{f}_{n}({\mb q_2})+
{\cal F}_n$. For these, the recurrence relations are $\delta \bar{\bar
f}_n({\mb q}_1, {\mb q}_2) = \bar{g}_0(\omega-n\Omega)\left[(n-2)g^2
\delta \bar{\bar f}_{n-1}({\mb q}_1, {\mb q}_2) +\delta \bar{\bar
f}_{n+1}({\mb q}_1, {\mb q}_2) \right] $ with the solution $ \delta
\bar{\bar f}_3({\mb q}_1,{\mb q}_2) = g^2 A_1(\omega-2\Omega)\delta
\bar{\bar f}_2({\mb q}_1,{\mb q}_2)$, from which we find $\bar{\bar
f}_3({\mb q}_1,{\mb q}_2) = g^2 A_1 f_2^{(2)}({\mb q}_1,{\mb q}_2) +
g^2 (A_2-A_1) \left[\delta\bar{f}_{2}({\mb q_1}) +
\delta\bar{f}_{2}({\mb q_2})\right]+ g^2\left[A_3 - A_1\right]{\cal
F}_2 $, i.e. only in terms of various partial averages of
$f_2^{(2)}({\mb q}_1,{\mb q}_2)$. Throughout we use the shorthand
notation
$$A_1 \equiv A_1(\omega- 2\Omega), A_2\equiv A_2(\omega - \Omega),
A_3\equiv A_3(\omega).$$

We now substitute $\bar{\bar f}_3({\mb q}_1,{\mb q}_2) $ in
Eq. (\ref{e22}), which gives:
\begin{eqnarray}
\nonumber & &f_2({\mb q}_1,{\mb q}_2) = g^2 G_0({\mb k}-{\mb q}_1 -
{\mb q}_2, \tilde{\omega}) \left[f_1^{(2)}({\mb q}_1) + f_1^{(2)}({\mb
q}_2)\right.\\
\label{f2} & &\left. + (A_2-A_1)  \left[\delta\bar{f}_{2}({\mb
  q_1}) + \delta\bar{f}_{2}({\mb q_2})\right]+\left[A_3 -
  A_1\right]{\cal F}_2\right].
\end{eqnarray}
We used the fact that $[G_0({\mb k}-{\mb q}_1 - {\mb q}_2,
  \omega-2\Omega)]^{-1} - g^2A_1 = [G_0({\mb k}-{\mb q}_1 - {\mb q}_2,
  \tilde{\omega})]^{-1}$, with $\tilde{\omega} = \omega - 2\Omega -
  g^2 A_1$.

Momentum averaging Eq. (\ref{f2}) over both momenta leads immediately
to ${\cal F}_2 = g^2 \bar{g}_0(\tilde{\omega})\left[2 {\cal F}_1 +
(A_3-A_1) {\cal F}_2\right]$, and therefore:
\begin{equation}
\EqLabel{F2} {\cal F}_2 = \frac{2g^2 \bar{g}_0(\tilde{\omega}){\cal
F}_1}{ 1- g^2 \bar{g}_0(\tilde{\omega})[A_3 - A_1]}.
\end{equation}
It follows that all higher order total averages ${\cal F}_n$ are
proportional to ${\cal F}_1=\Sigma$.

Since only $\bar{f}_2({\mb q}_1)$ is needed in Eq. (\ref{e31}), we can
obtain it by momentum averaging Eq. (\ref{f2}) over ${\mb q}_2$. This
leads directly to the closed system of coupled equations:
\begin{equation}
\nonumber 
f_1^{(2)}({\mb q_1})=G_0({\mb k} - {\mb q}_1,
\omega -\Omega)\left[g^2 + \delta \bar{f}_2({\mb q}_1) +{\cal F}_2\right],
\end{equation}
\begin{eqnarray}
\nonumber \delta \bar{f}_2({\mb q}_1) = g^2
\bar{g}_0(\tilde{\omega})\left[f_1^{(2)}({\mb q_1}) + (A_2-A_1)\delta
\bar{f}_2({\mb q}_1)- 2{\cal F}_1\right] && \\ \nonumber 
+{g^2\over N} \sum_{{\mb q}_2}^{} G_0({\mb k}-{\mb q}_1 - {\mb q}_2,
\tilde{\omega})\left[f_1^{(2)}({\mb q_2}) + (A_2-A_1)\delta
\bar{f}_2({\mb q}_2)\right]. &&
\end{eqnarray}
To solve it, we first rewrite these as a single equation in terms of
the unknowns $x_{\mb q} = f_1^{(2)}({\mb q}) + (A_2-A_1)\delta
\bar{f}_2({\mb q})$. This is obtained by adding the second equation to
$[A_2-A_1 + G_0({\mb k} - {\mb q}_1, \omega -\Omega)]$ times the first
equation. We introduce the short-hand notation:
$$ a_{ij}\equiv a_{ij}(\omega)=1 - g^2 \bar{g}_0(\tilde{\omega})[A_i -
A_j]
$$ and use the identities $[G_0({\mb k} - {\mb q}_1, \omega
-\Omega)]^{-1} - g^2 \bar{g}_0(\tilde{\omega})\left[1+ (A_2 - A_1)
(G_0({\mb k} - {\mb q}_1, \omega -\Omega))^{-1} \right]= [1- g^2
\bar{g}_0(\tilde{\omega})(A_2-A_1)] \left[G_0({\mb k} - {\mb
q}_1,\tilde{\tilde{\omega}})\right]^{-1}$, where
$$ \tilde{\tilde{\omega}}= \omega - \Omega -{g^2
\bar{g}_0(\tilde{\omega})\over a_{21}},
$$ and $G_0({\mb k} - {\mb
q}_1,\tilde{\tilde{\omega}})\left[1+(A_2-A_1) \left[G_0({\mb k} - {\mb
q}_1, \omega -\Omega)\right]^{-1}\right]= A_2 - A_1 + G_0({\mb k} -
{\mb q}_1,\tilde{\tilde{\omega}})/ a_{21}$. Both identities are based
on the definition $[G_0({\mb k}, \omega)]^{-1} = \omega -
\epsilon_{\mb k} + i \eta$.

With these, the final equation for $x_{\mb q}$ becomes:
\begin{eqnarray}
\nonumber a_{21} x_{\mb q} = G_0({\mb k} - {\mb
  q}_1,\tilde{\tilde{\omega}})\left[g^2 + {a_{21}-a_{31}\over
  a_{21}}{\cal F}_2\right] &&\\ \nonumber - a_{31}(A_2-A_1) {\cal
  F}_2+\left[A_2 - A_1 + {G_0({\mb k} - {\mb
  q}_1,\tilde{\tilde{\omega}})\over a_{21}}\right]&&\\ \nonumber
  \times{g^2\over N} \sum_{{\mb q}_2}^{} G_0({\mb k} -{\mb q} - {\mb
  q_2}, \tilde{\omega}) x_{{\mb q}_2}. &&
\end{eqnarray}

We now Fourier transform $x(i) = {1\over N} \sum_{\mb q}^{}
e^{i\mb{q}\cdot {\mb R}_i} x_{\mb q}$. From the definition of $x_{\mb
q}$ it follows immediately that $x(0) = {\cal F}_1= \Sigma$. Since
${\cal F}_2 \sim {\cal F}_1= x(0) $ [see Eq. (\ref{F2})], the
resulting system of inhomogeneous equations is:
$$ \sum_{j}^{} M_{ij}({\mb k},\omega) x(j) = e^{i\mb{k}\cdot {\mb
R}_i}g^2 G_0(-i, \tilde{\tilde{\omega}})
$$ where $G_0(i,\omega) = {1\over N} \sum_{\mb k}^{} e^{i\mb{k}\cdot
{\mb R}_i} G_0({\mb k}, \omega)$, and the matrix elements are:
\begin{eqnarray}
\label{moo} M_{00} = 1- g^2
\bar{g}_0(\tilde{\omega})\bar{g}_0(\tilde{\tilde{\omega}})\left({2\over
 a_{31}} - {1\over a_{21}}\right), &&\\
\label{mio} M_{i0} = - g^2 \bar{g}_0(\tilde{\omega})e^{i\mb{k}\cdot {\mb
 R}_i}G_0(-i, \tilde{\tilde{\omega}}) \left({2\over a_{31}} - {1\over
 a_{21}}\right) &&
\end{eqnarray}
for $i\ne 0$, and for both $i, j\ne 0$:
\begin{eqnarray}
\nonumber && M_{ij} = a_{21} \delta_{i,j} - g^2 e^{i\mb{k}\cdot {\mb
 R}_i}G_0(j,\tilde{\omega})\\
\label{mij} && \times \left[(A_2-A_1)\delta_{i,-j} +
 {G_0(-i-j,\tilde{\tilde{\omega}}) \over a_{21}}\right].
\end{eqnarray}
We truncate this system by keeping only sites ${\mb R}_i$ within a
certain distance from the origin (see discussion in main text) and
solve it numerically. The self-energy is
$$ \Sigma_{MA^{(2)}}({\mb k}, \omega) = x(0).
$$

\end{document}